\documentclass[twocolumn, jcp, superscriptaddress]{revtex4}
\usepackage{graphicx}
\usepackage[caption=false]{subfig}
\usepackage{mathtools}
\usepackage{algorithm}

\begin{document}

\title{Combining the Ensemble and Franck-Condon Approaches for Spectral Shapes of Molecules in Solution}
\author{T. J. Zuehlsdorff}
\email{tzuehlsdorff@ucmerced.edu}
\affiliation{School of Natural Sciences, University of California Merced, N. Lake Road, CA 95344, USA}
\author{C. M. Isborn}
\email{cisborn@ucmerced.edu}
\affiliation{School of Natural Sciences, University of California Merced, N. Lake Road, CA 95344, USA}

\begin{abstract}
The correct treatment of vibronic effects is vital for the modeling of absorption spectra of solvated dyes, as many prominent spectral features can often be ascribed to vibronic transitions. Vibronic spectra can be easily computed within the Franck-Condon approximation for small dyes in solution using an implicit solvent model. However, implicit solvent models neglect specific solute-solvent interactions and provide only an approximate treatment of solvent polarization effects on the electronic excited state. Furthermore, temperature-dependent solvent broadening effects are often only accounted for using a broadening parameter chosen to match experimental spectra. On the other hand, ensemble approaches provide a straightforward way of accounting for solute-solvent interactions and temperature-dependent broadening effects by computing vertical excitation energies obtained from an ensemble of solute-solvent conformations. However, ensemble approaches do not explicitly account for vibronic effects and thus often produce spectral shapes in poor agreement with experiment. We address these shortcomings by combining the vibronic fine structure of an excitation obtained in the Franck-Condon picture at zero temperature with vertical excitations computed for a room-temperature ensemble of solute-solvent configurations.  In this combined approach, all temperature-dependent broadening is therefore treated classically through the sampling of configurations, with quantum mechanical vibronic contributions included as a zero-temperature correction to each vertical transition. In our calculation of the vertical excitations, significant parts of the solvent environment are treated fully quantum mechanically to account for solute-solvent polarization and charge-transfer, whereas for the Franck-Condon calculations, a small amount of solvent is considered in order to capture explicit solvent effects on the vibronic shape function. We test the proposed method on Nile Red and the green fluorescent protein chromophore in polar and non-polar solvents. For systems with strong solute-solvent interaction, the approach yields a significant improvement over the ensemble approach, whereas for systems with weak to moderate solute-solvent interactions, both the shape and the width of the spectra are in excellent agreement with experiment. 
\end{abstract}

\date{\today}

\maketitle

\section{Introduction}
Accurate modeling of absorption spectra is key to interpreting experimental spectra, identifying molecular species, understanding the relative energetics of molecules in different environments, and predicting photochemistry. Many phenomena contributing to an absorption spectrum may need to be taken into account for accurate modeling, for example, the relative energies and intensities of various electronic transitions and their coupling to vibrational transitions, the configurations sampled by the molecule, and the finite lifetime of the excited state. For a molecule in solution, the absorption spectrum is broadened by the inhomogeneous environment provided by the various configurations of the solvent around the molecule; the configurations adopted by a molecule in solution might differ from those in vacuum.  Taking all of these spectral contributions into account is quite theoretically challenging since it requires accurate models for solute-solvent interactions, sampling of solute-solvent nuclear configurations, and accounting for vibrational and electronic transitions in a solvent environment. Often, the excitation energy of a molecule at its optimized geometry is simply computed with an implicit solvent model\cite{PCM1,PCM2,PCM3,PCM4}, ignoring explicit solute-solvent interactions, and this energy is then compared to the experimental absorption spectrum $\lambda_\textrm{\scriptsize{max}}$ value. This approach is commonly used to benchmark excited state methods (see, for example, discussions in \cite{Schreiber2008, Silva-Junior2010}), but the comparison of a single excitation energy to the experimental $\lambda_\textrm{\scriptsize{max}}$ can be problematic since many factors contributing to the spectral shape are left out of the theoretical model\cite{Lasorne2014}, including all vibronic and temperature effects. If comparison with or interpretation of experimental spectra is desired, then the goal should be to model the broadening of spectral lines to simulate a full absorption spectrum.

Being able to reproduce both the energy and the shape of the spectrum would be a rigorous test of a theoretical method, since it would require both accurate solute-solvent geometries, as well as accurate excited state methods for modeling the energy gap and intensity of transitions between electronic/vibrational states. Recent work\cite{photoactive_yellow_protein_isborn,baroni_spectral_warping,Alizarin_water,Nile_red_force field,protein_convergence1,protein_convergence2,FMO_danny,Joel_pct, Makenzie_pct} has shown that the explicit quantum mechanical (QM) treatment of significant parts of the solvent environment is essential in obtaining accurate and well-converged excitation energies in a variety of systems, ranging from small dyes in solution to pigment-protein complexes. Because sampling of the inhomogeneous solvent environment is required, and because it is likely necessary to include the solvent explicitly in the quantum mechanical (QM) electronic structure calculation to obtain proper solute-solvent polarization and charge-transfer, the excited state method must be computationally efficient enough to perform many hundreds of calculations on a large QM region (with hundreds of atoms). 

Time-dependent density functional theory (TDDFT)\cite{runge_gross,lr_tddft_original,casida_tddft} presents a good balance of accuracy and computational efficiency\cite{TDDFT_review}, so is often the method of choice for excited state calculations on larger systems (above 30 atoms or so). Although TDDFT has known problems modeling excited states in explicit solvent due to spuriously lowering the energy of solute-solvent charge-transfer transitions\cite{Bernasconi2003, Neugebauer2006}, these problems can be largely resolved by using a density functional with a large amount of exact exchange at long-range and using a classical solvent environment\cite{Lange2007, isborn_ct,danny_ct}, such as point charges or a polarizable continuum, to surround the QM solvent region. 

There are two dominant approaches for going beyond the computation of a single electronic excitation energy to compute a full absorption spectrum. The first way is through an ensemble approach, in which a Boltzmann sampling of nuclear configurations is obtained, often via a molecular dynamics (MD) ground state trajectory, and then excited state calculations are performed for each configuration\cite{explicit_solvent_excitation1,explicit_solvent_excitation2,explicit_solvent_excitation3,explicit_solvent_excitation4,explicit_solvent_excitation5,baroni_spectral_warping,Nile_red_force field, DeMitri2013}. Many configurations, often termed `snapshots' when obtained from an MD trajectory, are necessary for converging an absorption spectrum, so efficient excited state methods such as TDDFT are often used with this technique. The ensemble approach can easily take into account temperature effects and the molecule will naturally sample anharmonic regions of the potential energy surface according to the temperature of the system, making the approach suitable for flexible systems. Configurations can be obtained for molecules with their explicit solvent environment, which can then straight-forwardly be included in the QM excited state calculation at the same level of theory as the solute. The configurations adopted by the solute because of the presence of solvent will be included in the ensemble, and the excited states will also include differences in the transition dipole moment due to solute-solvent geometry. The ensemble approach therefore has many attractive qualities for modeling absorption spectra of molecules in solution; however, this method does not include effects that arise from the quantum nature of the nuclei, unless these are explicitly accounted for in the sampling method, such as by performing path integral molecular dynamics \cite{PIMD1,PIMD2}. Furthermore, there is generally no explicit accounting for nuclear wave functions on the excited state potential energy surface, and so no contributions from simultaneous excitations of electronic and vibrational states are considered, which can result in poor line shapes in comparison to experiment\cite{Furche_vibronic}. The ensemble approach is thus appropriate for a system behaving classically, such as a very flexible molecule at high temperature; systems such as this may have an absorption spectrum dominated by solvent induced inhomogeneous broadening, with no vibronic contributions. 

The other method for simulating absorption spectra is with a vibronic approach\cite{vibronic_harmonic1,vibronic_harmonic2,vibronic_prescreening1,vibronic_prescreening2,Vibronic_duschinsky}. This method accounts for the vibrational transitions that accompany electronic transitions, including the vibrational wave functions of the ground and excited electronic state that lead to the vibronic fine structure of the spectrum. The Franck-Condon principle is applied by assuming a vertical transition from the ground electronic state, and then the intensity of the vibronic transitions can be computed from the overlap of the ground and excited state vibrational nuclear wave functions obtained from the shape of the ground and excited state potential energy surface (PES). Most Franck-Condon vibronic approaches make use of a harmonic approximation\cite{vibronic_harmonic1,vibronic_harmonic2} to the shape of the ground and excited state PESs in order to simplify computation of the nuclear wave functions, but there are some techniques that include anharmonic effects, resulting in more accurate vibronic fine structure\cite{Vibronic_frozen_solute,Santoro_GFP_solution,Santoro_mixed_quantum_classical}. Accounting for the temperature of a system can lead to population beyond the ground vibrational state\cite{FCclasses1,FCclasses2}, but this effect is usually minimal beyond soft, anharmonic vibrational modes. Often, with this approach the absorption spectrum is computed at 0~K, so that only one vibrational wave function ($v$=0) is considered for the ground electronic state. The Franck-Condon approach is appropriate for a rigid molecule at low temperature since such systems have absorption spectra dominated by vibronic fine structure. 

Spectra computed with the vibronic approach usually do not account for solvent effects beyond including implicit solvent in the ground and excited state optimized geometries used in the Franck-Condon calculation. However, the solvent environment will often lead to much broader absorption spectra than those obtained in vacuum. In general, the spectral broadening due to the inhomogeneous solvent environment is assumed to be Gaussian in nature and a Gaussian function is used to broaden vibronic peaks.  The standard deviation of the Gaussian function can either be estimated in implicit solvent calculations following Marcus theory\cite{marcus_theory}, or from considering an explicit sampling of solvent conformations\cite{Vibronic_frozen_solute,Santoro_vibronic}. Estimating the solvent broadening from explicit solvent configurations generally assumes a complete decoupling of solute and solvent degrees of freedom\cite{Santoro_vibronic, Vibronic_frozen_solute}, such that the solvent configurations are sampled from MD around a solute frozen in its optimized ground state geometry. Under the approximation of a decoupling of solute and solvent degrees of freedom, the Gaussian broadening width can then be extracted from the spread TDDFT excitation energies computed from the ensemble of solvent configurations around the frozen solute. Estimating the broadening from explicit solvent conformations generated using MD is appealing since it could potentially account for both the effects of temperature and explicit solute-solvent interactions contributing to inhomogeneous broadening. Taking solvent conformations around a solute frozen in its ground state structure, as suggested in \cite{Vibronic_frozen_solute,Santoro_vibronic} has the advantage that the system in question can be strictly separated into classical and quantum degrees of freedom. However, this strict separation of solute and solvent degrees of freedom could lead to a sampling of incorrect solute-solvent configurations. Small dyes can be stabilized in twisted conformations due to direct solute-solvent interactions \cite{Santoro_GFP_solution} and the first solvation shell is very dependent on the dynamics of the chromophore. Furthermore, treating all solute degrees of freedom as harmonic in order to facilitate the computation of nuclear wave functions likely reaches its limit for semi-flexible systems, where low frequency modes that dominate the motion of the system at low temperature shows considerable anharmonicity. Using this frozen solute-dynamic solvent technique led to computed absorption spectra that were too narrow compared to experiment, and that had some disagreement with spectral shape\cite{Santoro_vibronic,Santoro_GFP_solution}. 

The absorption spectra for many dyes in solution have large contributions from vibronic broadening as well as inhomogeneous broadening from solute-solvent interactions. However, these spectral broadening effects are difficult to model together. In this work, we combine the Franck-Condon vibronic approach with the ensemble configuration sampling approach for modeling the band shape of dyes in solution. The Franck-Condon spectrum is computed at the zero temperature limit for the dye in a small amount of frozen explicit solvent to capture the quantum mechanical vibronic effects present in the fast vibrational modes of the dye. All broadening due to temperature effects and to explicit solute-solvent interactions, as well as from sampling anharmonic regions of the PES, is treated classically. Spectral broadening from these effects is simulated by TDDFT calculations performed on an ensemble of configurations obtained from an MD trajectory. The spectral lines obtained from this Boltzmann ensemble of configurations are then broadened with the zero-temperature Franck-Condon spectrum. The approach makes a variety of assumptions, which we detail in the theoretical background section, but the most important of these assumptions are: 1) that although the configurations of the system are sampled from anharmonic regions of the PES, the Franck-Condon spectrum obtained at the optimized geometry is still a good approximation to the vibronic spectrum at that configuration and that the ground state vibrational mode is predominantly populated, 2) that the ensemble approach is capturing the appropriate temperature dependence of the system and that we can neglect a quantum treatment of temperature in the vibronic spectrum, and 3) that the solvent vibrational modes do not directly couple to the solute vibrational modes, so can be neglected in the calculation of the Franck-Condon shape function. Furthermore, we assume that we can neglect broadening effects due to the finite lifetime of the transition, since these should be smaller than the vibronic and solvent-induced broadening.

\begin{figure}
\centering
\subfloat[GFP anion \label{subfig:gfp_anion_struct}]{\resizebox{0.14\textwidth}{!}{\includegraphics{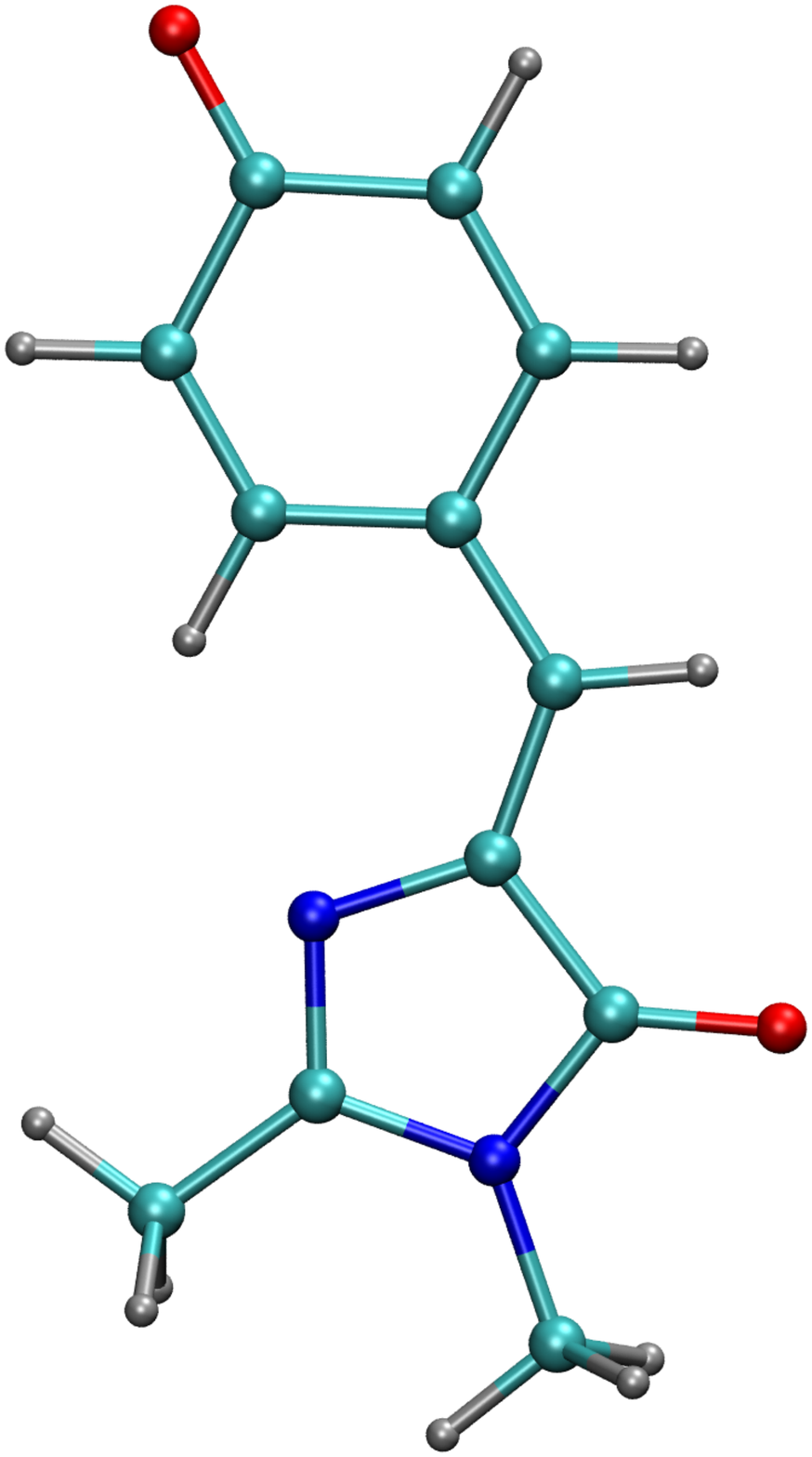}}}
\subfloat[GFP neutral+ \label{subfig:gfp_neutral_plus_struct}]{\resizebox{0.16\textwidth}{!}
{\includegraphics{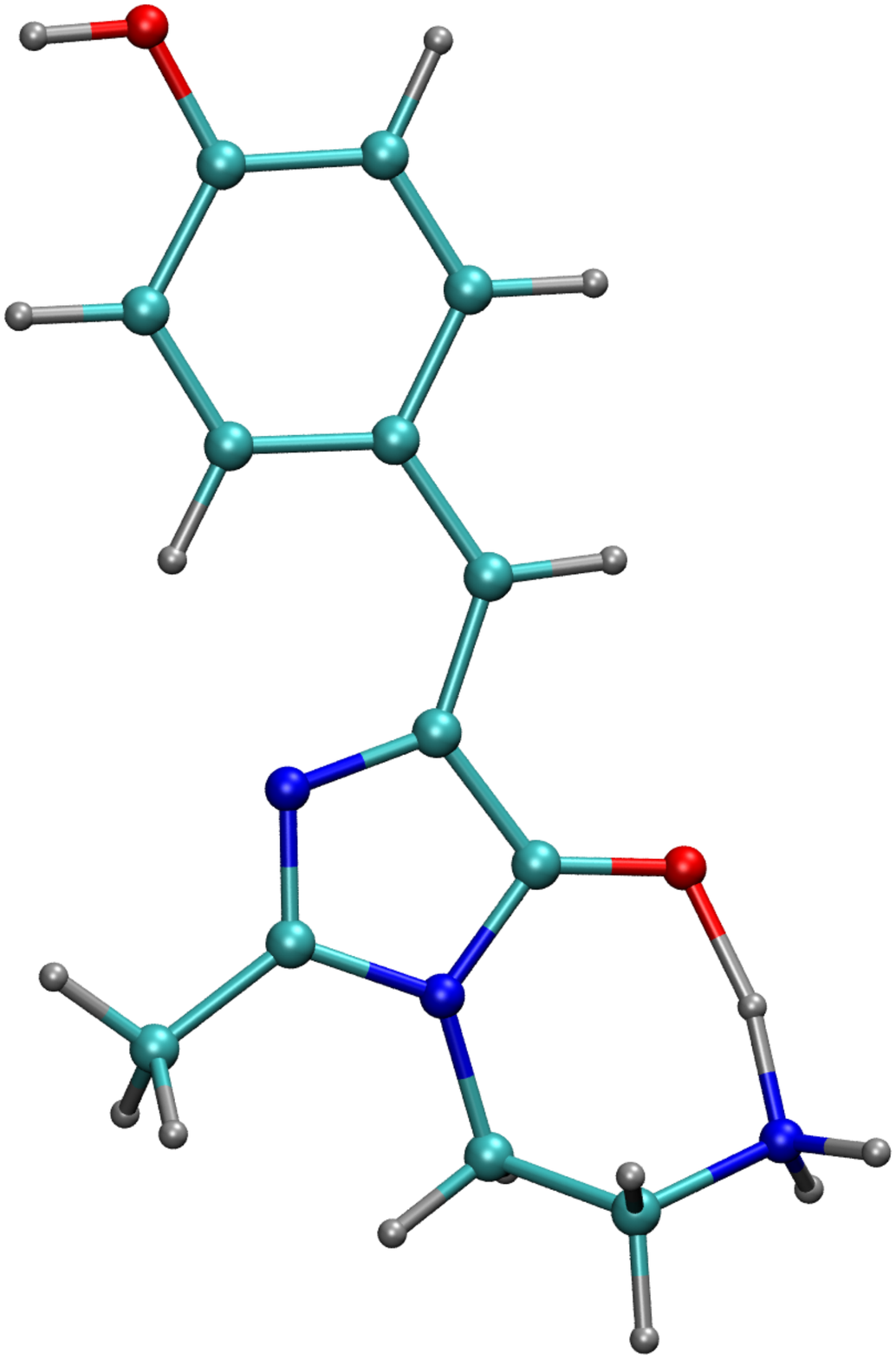}}} 
\subfloat[Nile Red \label{subfig:nile_red_struct}]{\resizebox{0.19\textwidth}{!}
{\includegraphics{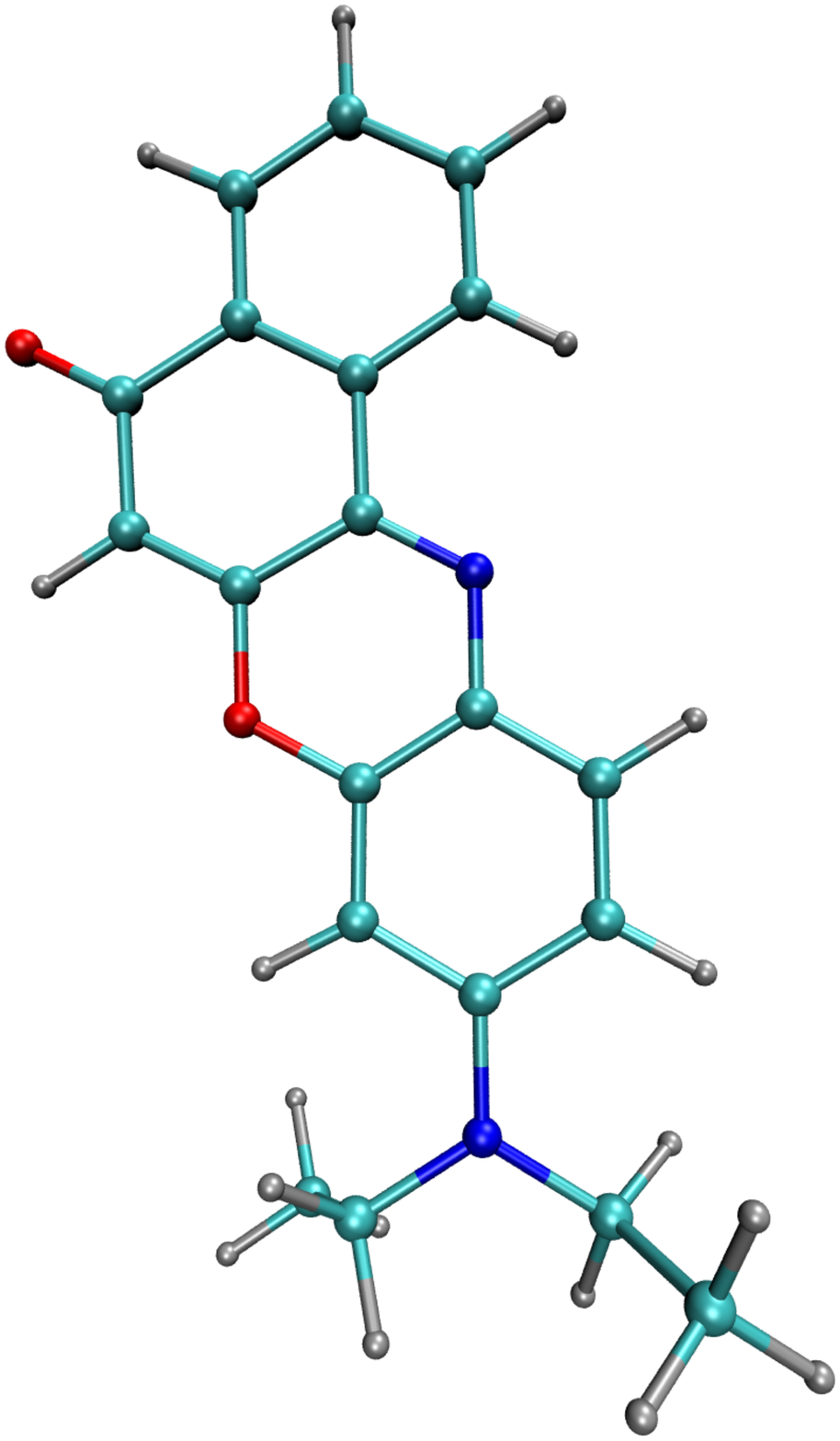}}}
\caption{Ball and stick representation of the ground state structure of the three small dyes that form the focus of this work. This figure was created using VMD\cite{VMD}.}
\label{fig:structures_dyes}
\end{figure}

In this article we outline the theory of simulating spectral shapes with the combined ensemble and Franck-Condon approach and then we test this approach on three dyes in different solvents. The tests are on the chromophore of green fluorescent protein (GFP) as an anion in water\cite{GFP_anion_experiment}, a positively charged analogue of the neutral GFP chromophore in methanol\cite{GFP_neutral_plus_experiment}, and the Nile Red chromophore in acetone, benzene, and cyclohexane\cite{Nile_red_experiment_spectrum} (see Fig. \ref{fig:structures_dyes} for the molecular structures of all dyes studied in this work). We choose two non-polar solvents and a polar solvent in order to attempt to reproduce the significant changes in spectral shape with increasing solvent polarity observed in Nile Red\cite{Nile_red_experiment_spectrum}. The GFP anion and neutral+ variant are simulated in water and methanol, respectively, which allows us to assess the performance of the combined ensemble and Franck-Condon approach for systems with strong solute-solvent interactions from hydrogen bonds. The absorption spectrum of GFP in solution has been modeled with a variety of approaches, but it has proven challenging to theory to correctly simulate the spectrum from first principles without using any phenomenological broadening parameters\cite{Huang2012, Santoro_GFP_solution, Zutterman2017}. Our combined approach yields significant improvement in agreement with the experimental spectral shape compared to the ensemble approach alone, and appears to be a promising technique for modeling the absorption spectrum shape for dyes in solution, in which both vibronic and inhomogeneous solvent broadening are important.

\section{Theoretical Background}
\subsection{Vertical excitation energies}
The most straightforward approximation to computing absorption spectra of solvated dyes is to ignore dynamic effects and the quantum nature of the nuclei entirely. The excited states of the electronic system then depend parametrically on the nuclear positions, but any coupling between the electronic excitation and the nuclear degrees of freedom of the system, such as the simultaneous excitation of electronic and vibrational quantum states, is ignored. The excited states computed in this fashion are referred to as the bare vertical excitations of the system in this work. For a single bare vertical excitation, often the geometry of the dye is optimized, and then an electronic excited state calculation is performed at that geometry. Although this approach provides only a vertical excitation energy with no notion of spectral shape, the bare vertical excitation energies are often convoluted with a Gaussian function and compared with experimental spectra. The absorption cross section (in atomic units) of a system with a single electronic transition $i\rightarrow f$ of interest can then be written as:
\begin{eqnarray} \nonumber
\sigma^{\textrm{\scriptsize{vert}}}(\omega)&=&\frac{4\pi^2\omega_{if}(\textbf{R}_0^{\textrm{\scriptsize{GS}}})}{3c} \times \\
&&\left|\boldsymbol{\mu}_{if}(\textbf{R}_0^{\textrm{\scriptsize{GS}}}) \right|^2 \mathcal{N}\left(\omega_{if}(\textbf{R}_0^{\textrm{\scriptsize{GS}}}), \sigma(T)^2 \right).
\label{eqn:pcm_spectrum}
\end{eqnarray}
Here, $\omega_{if}$ denotes the excitation energy of the electronic transition going from initial state $i$ to final state $f$, $\boldsymbol{\mu}_{if}$ is the electronic transition dipole moment of the transition, and $\textbf{R}$ is a collective variable of the nuclear positions of the chromophore atoms, such that $\textbf{R}_0^{\textrm{\scriptsize{GS}}}$ is the optimized ground state structure on the ground state potential energy surface. The symbol $\mathcal{N}$ denotes a Gaussian function with mean $\omega_{if}$ and standard deviation $\sigma(T)$, where $\sigma(T)$ is taken to be a broadening parameter accounting for all temperature-dependent broadening effects ignored in the calculation, both due to fluctuations in the environment and vibrations of the dye itself. In practice, $\sigma(T)$ is often chosen in a purely phenomenological way in order to reproduce the width of experimental spectra.

Having made the choice of ignoring all dynamic effects and the quantum nature of the nuclei in the computation of the absorption spectra, one has to consider how to represent the solvent environment of the chromophore. The two basic choices that can be made are between explicit solvent representations\cite{explicit_solvent_excitation1,explicit_solvent_excitation2,explicit_solvent_excitation3,explicit_solvent_excitation4,explicit_solvent_excitation5,photoactive_yellow_protein_isborn,baroni_spectral_warping,Alizarin_water,Nile_red_force field} and polarizable continuum models (PCMs)\cite{PCM1,PCM2,PCM3,PCM4}. 

In a PCM representation, the chromophore is placed in a dielectric medium that represents the average electrostatic screening of the solvent environment. Although a wide variety of different implicit solvation approaches have been developed\cite{PCM1,PCM2}, their common feature is that they represent a computationally efficient way of treating average solvation effects, thus allowing for the use of accurate \emph{ab-initio} quantum chemistry methods to solve for the electronic excitations of small to medium-sized dyes in solution. Absorption spectra in solution are often modeled within a PCM by considering that electronic excitations occur on a much faster timescale than the movement of the solvent atoms. The excited states are then computed in a non-equilibrium solvation model, where it is assumed that only the fast electronic degrees of freedom of the solvent can react to the instantaneous change in density during the excitation\cite{solvent_lr,solvent_ss,solvent_ss2}. The computed excitation energy and transition dipole moment for a chromophore in its ground state structure can then be directly used in Eqn. \ref{eqn:pcm_spectrum} to yield an approximate absorption spectrum including average polarization effects by the solvent environment. However, in systems with significant solute-solvent interactions, pure PCM treatments can fail in correctly recovering absorption properties and solvatochromic shifts\cite{Alizarin_water,Nile_red_force field,Joel_pct, Makenzie_pct}. 

Combining Eqn. \ref{eqn:pcm_spectrum} with a PCM includes the average solvent screening in the electronic excitation of the chromophore, but it cannot account for direct solute-solvent interactions and does not include inhomogeneous broadening and temperature effects beyond the ad hoc parameter $\sigma(T)$. Thus, if one is interested in the shape of the absorption spectrum rather than just the position of the absorption maximum, an explicit representation of the solvent degrees of freedom has to be added to the model. A popular way of accounting for the effect of the explicit solvent environment on the absorption spectrum is the ensemble approach.

\subsection{The ensemble approach}
In the ensemble approach, the nuclei are still treated as purely classical particles, and it is straightforward to include an explicit representation of both the solute and solvent degrees of freedom through a conformational integral: 
\begin{eqnarray} \nonumber
\sigma^{\textrm{\scriptsize{vert}}}(\omega)&=&\frac{4\pi^2\omega}{3c} \int \textrm{d}\textbf{R}\,\rho^{\textrm{\scriptsize{GS}}}_{\textrm{\scriptsize{cl}}}(\textbf{R},T) \times \\
&& \left|\boldsymbol{\mu}_{if}(\textbf{R}) \right|^2 \delta\left(\omega-\omega_{if}(\textbf{R})\right)
\label{eqn:vert_conf_integral}
\end{eqnarray}
Here, $\textbf{R}$ is a collective variable for all solute and solvent degrees of freedom and $\rho^{\textrm{\scriptsize{GS}}}_{\textrm{\scriptsize{cl}}}(\textbf{R},T)$ denotes a classical probability distribution function for a given solute-solvent conformation $\textbf{R}$, with the electronic system in its ground state, at temperature $T$. In practice, the conformational integral of Eqn. \ref{eqn:vert_conf_integral} can be sampled using molecular dynamics (MD). Extracting $N_\textrm{\scriptsize{frames}}$ uncorrelated snapshots of solute-solvent conformations from a sufficiently long MD trajectory sampling the ground state potential energy surface of the system, Eqn. \ref{eqn:vert_conf_integral} can be approximated as
\begin{equation}
\sigma^{\textrm{\scriptsize{vert}}}_{\textrm{\scriptsize{MD}}}(\omega)=\frac{1}{N_\textrm{\scriptsize{frames}}}\frac{2\pi^2}{c}\sum_j^{N_\textrm{\scriptsize{frames}}} f_{if}(\textbf{R}_j) \,\mathcal{N}\left(\omega_{if}(\textbf{R}_j), \sigma^2 \right)
\label{eqn:vert_MD}
\end{equation}
where $\{\textbf{R}_j\}$ is the set of uncorrelated snapshots extracted from the MD trajectory and $f_{if}(\textbf{R}_j)=\frac{2}{3}\omega_{if}(\textbf{R}_j) \left|\boldsymbol{\mu}_{if}(\textbf{R}_j) \right|^2$ denotes the oscillator strength of the electronic transition $i\rightarrow f$ for a given conformation $\textbf{R}_j$. Note that, in this case, the Gaussian function $\mathcal{N}$ is simply introduced to guarantee a smooth spectrum for finite sampling. Thus $\sigma$ is no longer related to any physical effects like inhomogeneous solvent broadening, but is rather just a small numerical convergence parameter such that $\sigma\rightarrow0$ as $N_\textrm{\scriptsize{frames}}\rightarrow \infty$.

Approximating the absorption spectrum of a solvated dye in terms of a sampling over conformations comes with a number of advantages. Both solute-solvent interactions and temperature broadening effects are accounted for in a straightforward fashion. Furthermore, the shape of the spectrum is no longer limited to be a Gaussian distribution as in Eqn. \ref{eqn:pcm_spectrum}, which is not necessarily expected in nature for a system with strong solute-solvent interactions. Given that the MD naturally includes sampling of anharmonic parts of the ground state potential energy surface, as well as captures specific solute-solvent interactions, the distribution of vertical excitation energies is generally not expected to be Gaussian. Additionally, parts of the solvent environment can be treated explicitly quantum mechanically in the computation of the excitation energies $\omega_{if}(\textbf{R}_j)$ and transition dipole moments $\boldsymbol{\mu}_{if}(\textbf{R}_j)$, allowing for a consistent treatment of polarization effects in the ground and excited state at the QM level.  

Although computing approximate absorption spectra of solvated systems according to Eqn. \ref{eqn:vert_MD} using a QM treatment of significant parts of the solvent environment is advantageous for the reasons specified above, it comes at a significant increase in computational cost compared with simple PCM treatments. Furthermore, to obtain a sufficiently accurate approximation of the the conformational integral, $N_\textrm{\scriptsize{frames}}$ individual excited state calculations have to be carried out, where $N_\textrm{\scriptsize{frames}}\approx\mathcal{O}(100)-\mathcal{O}(1000)$\cite{photoactive_yellow_protein_isborn}. This increased computational cost means that excitation energies and transition dipole moments are typically computed within time-dependent density functional theory (TDDFT)\cite{runge_gross,lr_tddft_original,casida_tddft}, due to its good compromise between computational complexity and achievable accuracy\cite{TDDFT_review}. 

It should be noted that although Eqn. \ref{eqn:vert_MD} accounts for temperature-dependent inhomogeneous spectral broadening in an intuitive manner, it still represents a fully classical treatment of the nuclear degrees of freedom of the system, and therefore does not allow for the simultaneous excitation of electronic and vibrational states. For many small chromophores in solution, the optical absorption spectra are known to have a significant contribution from vibronic states and a simulation of the spectrum in the form of Eqn. \ref{eqn:vert_MD} often fails in reproducing the fine structure of the spectral shape, as well as its width\cite{Furche_vibronic,photoactive_yellow_protein_isborn}. 

\subsection{Franck-Condon spectra}
In order to treat vibronic effects in the absorption spectra of small dyes, it is necessary to account for the quantum nature of the nuclei. Working under the Born-Oppenheimer approximation, such that the total wave function of the system is separable into a nuclear and an electronic part, the absorption cross section for a given excitation from electronic state $i$ to electronic state $f$ can be written as\cite{FCclasses1,FCclasses2,Furche_vibronic}:
\begin{eqnarray} \nonumber
\sigma^{\textrm{\scriptsize{vib}}}(\omega)=\frac{4\pi^2\omega}{3c}\sum_{v_{i}}\rho(v_i,T) \times \\
\sum_{v_f} \left|\langle\phi_{v_i}| \boldsymbol{\mu}_{if}|\phi_{v_f} \rangle \right|^2 \delta(E_{v_f}-E_{v_i}-\omega)
\end{eqnarray} 
Here, $\phi_{v_i}$ and $\phi_{v_f}$ denote nuclear wave functions of the Born-Oppenheimer PES of electronic state $i$ and $f$ respectively and $\boldsymbol{\mu}_{if}$ is the electronic transition dipole moment given by $\boldsymbol{\mu}_{if}=-\langle \chi_i|\textbf{r} |\chi_f\rangle$ for electronic wave functions $|\chi_i\rangle$ and $|\chi_f\rangle$. $E_{v_f}$ denotes the total energy of the system in electronic state $f$ and vibrational mode $v_f$ and $\rho(v_i,T)$ is the Boltzmann population of a given initial vibrational state $|\phi_{v_i}\rangle$ at temperature $T$. 

The inner product $\langle\phi_{v_i}| \boldsymbol{\mu}_{if}|\phi_{v_f} \rangle$ cannot be trivially computed since $\boldsymbol{\mu}_{if}$ is dependent on nuclear positions through the parametrical dependence of the electronic wave functions on atomic positions. Treating $\boldsymbol{\mu}_{if}$ as a constant for the purpose of evaluating the inner product over nuclear wave functions yields the Franck-Condon approximation, such that the spectrum can be approximated as
\begin{eqnarray} \nonumber
\sigma^{\textrm{\scriptsize{vib}}}_{\textrm{\scriptsize{FC}}}(\omega)=\frac{4\pi^2\omega}{3c}|\boldsymbol{\mu}_{if}|^2\sum_{v_{i}}\rho(v_i,T) \times \\
\sum_{v_f} \left|\langle\phi_{v_i}| \phi_{v_f} \rangle \right|^2 \delta(E_{v_f}-E_{v_i}-\omega)
\label{eqn:abs_fc}
\end{eqnarray}
In this approximation, the absorption spectrum for a single given electronic transition is defined by a series of distinct vibronic peaks corresponding to excitations between different vibrational states, with the peak intensity given by the Franck-Condon overlaps $\langle\phi_{v_i}| \phi_{v_f} \rangle$. 

The expression in Eqn. \ref{eqn:abs_fc} can be evaluated directly under a number of further approximations. Approximating the Born-Oppenheimer PESs as harmonic around their respective minima\cite{vibronic_harmonic1,vibronic_harmonic2}, the vibrational states $\phi_{v_i}$ and $\phi_{v_f}$ can be computed from the Hessian at the two minima. The normal coordinates of the initial and final states are assumed to be related by the linear Duschinsky transformation\cite{Vibronic_duschinsky} and the Franck-Condon integrals can be directly computed by making use of powerful prescreening techniques to reduce the computational cost\cite{vibronic_prescreening1,vibronic_prescreening2,FCclasses1,FCclasses2}. Alternatively, Eqn. \ref{eqn:abs_fc} can also be formulated in the time domain to avoid the costly computation of individual Franck-Condon overlaps\cite{Furche_vibronic,vibronic_time_domain1,vibronic_time_domain2,vibronic_time_domain3}. In general, Eqn. \ref{eqn:abs_fc} is computationally expensive to evaluate due to the double sum over vibrational modes. However, in many systems of interest, the ground state vibrational mode is initially predominantly occupied at room temperature. Thus in many situations, we can make the approximation that $\sigma^{\textrm{\scriptsize{vib}}}_{\textrm{\scriptsize{FC}}}(\omega)\approx \lim_{T\rightarrow0}\sigma^{\textrm{\scriptsize{vib}}}_{\textrm{\scriptsize{FC}}}(\omega)$, turning the double sum over vibrational modes into a single sum over vibrational modes of the final state only. 

When considering a solvated chromophore, solvent effects can be straightforwardly introduced as was done in Eqn. \ref{eqn:pcm_spectrum} for the bare vertical excitations. To do so, the chromophore is placed in implicit solvent and the vibrational normal modes as well as energies are computed within the PCM. As in  Eqn. \ref{eqn:pcm_spectrum}, any inhomogeneous solvent broadening effects are assumed to be Gaussian in nature and are modeled through the parameter $\sigma(T)$. The absorption spectrum can then be written as
\begin{eqnarray} \nonumber
\sigma^{\textrm{\scriptsize{vib}}}_{\textrm{\scriptsize{FC}}}(\omega)=\frac{4\pi^2}{3c}|\boldsymbol{\mu}_{if}|^2\sum_{v_{i}}\rho(v_i,T) \times \\ 
\sum_{v_f} \left[ E_{v_f}-E_{v_i}\right]\left|\langle\phi_{v_i}| \phi_{v_f} \rangle \right|^2 \mathcal{N}\left(\left[ E_{v_f}-E_{v_i}\right], \sigma(T)^2 \right). 
\label{eqn:abs_fc_solv}
\end{eqnarray}

As with Eqn. \ref{eqn:pcm_spectrum}, the parameter $\sigma(T)$ is often chosen in an ad hoc way in order to yield spectra that are in good agreement with experimental data. However, to have a more predictive method of computing absorption spectra in solvated systems, $\sigma(T)$ should ideally be estimated from first principles. This can be achieved in a number of different ways. In a simple PCM treatment, $\sigma(T)$ can be related to the difference in excitation energy of the vertical transition as calculated in non-equilibrium solvation and equilibrium solvation following Marcus theory\cite{marcus_theory}. Alternatively, $\sigma(T)$ can be estimated in an explicit solvent treatment by assuming that the solute and solvent degrees of freedom are strictly independent\cite{Vibronic_frozen_solute,Santoro_vibronic}. Treating the solute degrees of freedom quantum mechanically through Eqn. \ref{eqn:abs_fc_solv} while the solvent degrees of freedom are considered purely classically, $\sigma(T)$ can be estimated from the spread of vertical excitation energies computed for a representative set of solvent conformations around the solute frozen in its ground state structure. 

Estimating $\sigma(T)$ through MD sampling of solvent conformations or PCM methods around the frozen solute can generate spectra that are too narrow compared to experimental absorption spectra\cite{Santoro_vibronic,Santoro_GFP_solution}. One approach to overcome these issues is to split the vibrations of the system into a number of low frequency modes that are accounted for purely classically in order to capture anharmonicity effects and conformational changes of the solute and high frequency modes that are accounted for quantum-mechanically in the harmonic approximation. This approach has been shown to yield very accurate results for the spectral shapes of oligothiophenes\cite{Santoro_mixed_quantum_classical}, but relies on chemical intuition in order to identify which motions of the system should be considered classically. 

\subsection{The combined ensemble plus vibronic broadening approach}
We propose that a more accurate way to simulate the absorption spectrum of flexible dyes in solution is to use a combined ensemble plus vibronic broadening approach.  First, for a given temperature, an ensemble of solute-solvent configurations would be obtained with a fully flexible solute. This would allow sampling of solute-solvent degrees of freedom missing in the frozen solute approach, such as twisted conformations of the solute along with the corresponding solvent environment. Next, the ensemble approach can be rigorously extended to include vibronic effects by computing a Franck-Condon spectrum for each snapshot in the ensemble. Following the main assumptions considering solvent-relaxation time-scales made in implicit solvent models, the Franck-Condon spectrum of a given MD snapshot can be computed by keeping the solvent atoms fixed and optimizing the dye structure in the ground and excited state in the frozen solvent pocket. The frozen solvent pocket contains the configurations of the solvent at the desired temperature, whereas the Franck-Condon spectrum contains the temperature effects of the solute within that solvent configuration.  The finite temperature Franck-Condon spectra computed this way can then be averaged over, yielding spectra that rigorously account for both direct solute-solvent interactions and anharmonic low frequency modes through the unconstrained MD sampling. However, this proposed combined approach is generally not computationally feasible, as it requires a full ground and excited state frequency calculation for $N_\textrm{\scriptsize{frames}}$ snapshots, in which large parts of the solvent environment are treated quantum mechanically. 

For the purpose of this work, rather than modeling the spectrum according to Eqn. \ref{eqn:abs_fc_solv} with a $\sigma(T)$ estimated by separating solute and solvent degrees of freedom or by computing a vibronic spectrum for each MD snapshot, we follow a different route. The aim is to use a zero temperature vibronic shape function that accounts for the overlap between the vibrational wave function of the electronic ground state and multiple vibrational modes on the excited state PES, while treating temperature fluctuations and solvent-induced inhomogeneous broadening fully classically through the ensemble approach.

\begin{figure}
\centering
\includegraphics[width=0.45\textwidth]{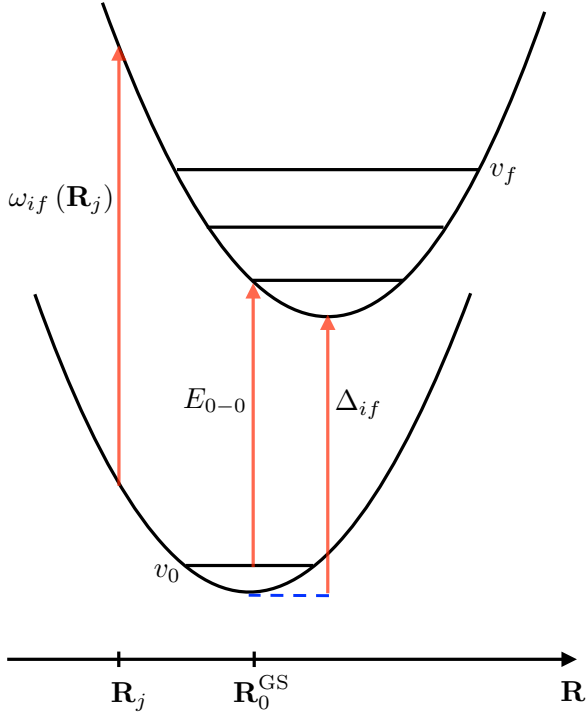}
\caption{Schematic showing the difference between the vertical excitation energy $\omega_{if}(\textbf{R}_j)$, the vibronic 0-0 transition energy $E_{0-0}$ and the adiabatic excitation energy $\Delta_{if}$ for a simple one-dimensional system. }
\label{fig:vibronic_schematic}
\end{figure}

As a first approximation, we take the limit of Eqn. \ref{eqn:abs_fc_solv} as $T\rightarrow 0$, such that we can write the bare Franck-Condon spectrum excluding any temperature effects as:
\begin{eqnarray} \nonumber
\sigma^{\textrm{\scriptsize{vib}}}_{\textrm{\scriptsize{FC}}}\left[T\rightarrow0\right](\omega;\sigma)=\frac{4\pi^2}{3c}|\boldsymbol{\mu}_{if}|^2 \times \\
\sum_{v_f} \left[E_{v_f}-E_{v^0_i}\right] \left|\langle\phi_{v^0_i}| \phi_{v_f} \rangle \right|^2 \mathcal{N}\left( \left[E_{v_f}-E_{v^0_i}\right],\sigma\right). 
\label{eqn:abs_fc_solv_0K}
\end{eqnarray}
We then reintroduce temperature in terms of purely classical fluctuations. To do so, we rewrite the average excitation energy $E_{v_f}-E_{v^0_i}$ for a given Franck-Condon transition as
\begin{eqnarray} \nonumber
E_{v_f}-E_{v^0_i}&=&E^{\textrm{\scriptsize{vib}}}_{0_i\rightarrow v_f}+\Delta_{if} \\ \nonumber
&&+\left[ \int \textrm{d}\textbf{R}\,\rho^{\textrm{\scriptsize{GS}}}_{\textrm{\scriptsize{cl}}}(\textbf{R},T)\,\omega_{if}(\textbf{R})-\omega_{if}(\textbf{R}_0^{\textrm{\scriptsize{GS}}})\right] \\
&=&E^{\textrm{\scriptsize{vib}}}_{0_i\rightarrow v_f}+\Delta_{if}+E^{\textrm{\scriptsize{av}}}_\textrm{\scriptsize{fluct}}(T)
\end{eqnarray}
where $E^{\textrm{\scriptsize{vib}}}_{0_i\rightarrow v_f}$ denotes the purely vibrational contribution to the vibronic energy in transitioning from the ground state vibrational mode of the ground state PES to the vibrational mode $v_f$ of the excited state PES, $\Delta_{if}$ is the adiabatic excitation energy while $E^{\textrm{\scriptsize{av}}}_\textrm{\scriptsize{fluct}}(T)$ is the average energy due to a purely classical treatment of temperature fluctuations (see Fig. \ref{fig:vibronic_schematic} for a schematic showing the relationship between $E_{0-0}$, $\Delta_{if}$ and $\omega_{if}(\textbf{R}_j)$). 

We next propose that the temperature-induced fluctuations of the system can be included in the absorption spectrum by conformational sampling on the same lines as in Eqn. \ref{eqn:vert_MD}:

\begin{eqnarray} \nonumber
\sigma^{\textrm{\scriptsize{vib}}}_{\textrm{\scriptsize{MD}}}(\omega;T)=\frac{4\pi^2}{3cN_{\textrm{\scriptsize{frames}}}} \sum^{N_\textrm{\scriptsize{frames}}}_j\sum_{v_f}\left|\boldsymbol{\mu}_{if} \left(\textbf{R}_0^{\textrm{\scriptsize{GS}}} \right)\right|^2\times  \\ \nonumber 
\left[E^{\textrm{\scriptsize{vib}}}_{0_i\rightarrow v_f}+\Delta_{if}+\omega_{if}(\textbf{R}_j)-\omega_{if}(\textbf{R}_0^{\textrm{\scriptsize{GS}}})\right] \left|\langle\phi_{v^0_i}| \phi_{v_f} \rangle \right|^2\times \\ \nonumber
\mathcal{N}\left(\left[E^{\textrm{\scriptsize{vib}}}_{0_i\rightarrow v_f}+\Delta_{if}+\omega_{if}(\textbf{R}_j)-\omega_{if}(\textbf{R}_0^{\textrm{\scriptsize{GS}}})\right], \sigma^2 \right) \\ \nonumber
\approx \frac{1}{N_{\textrm{\scriptsize{frames}}}} \sum^{N_\textrm{\scriptsize{frames}}}_j \frac{f_{if}\left(\textbf{R}_j\right)}{f_{if}\left(\textbf{R}_0^{\textrm{\scriptsize{GS}}}\right)}\times\\ 
\sigma^{\textrm{\scriptsize{vib}}}_{\textrm{\scriptsize{FC}}}\left[T\rightarrow0\right]\left(\omega-\omega_{if}(\textbf{R}_j)+\omega_{if}(\textbf{R}_0^{\textrm{\scriptsize{GS}}}); \sigma\right)
\label{eqn:vib_broadening}
\end{eqnarray}
where in the last line we have again introduced the electronic oscillator strength $f_{if}$ of transition $i\rightarrow f$ and $\sigma$ denotes a small numerical convergence parameter accounting for finite sampling. 
The expression at this point includes nuclear degrees of freedom of the solute both from temperature dependent vibrations that are sampled from the MD and from a zero-temperature ground state vibrational wave function obtained from a Franck-Condon calculation. However, much of the spread in vertical excitation energies is due to specific solute-solvent configurations, rather than solute-only vibrations. The orientation of solvent around a solute, and the resulting inhomogeneous broadening, is likely different for a flexible solute compared to a frozen solute. The advantage of the combined approach is that we retain a full coupling of the solute and solvent degrees of freedom during the MD sampling. Note that the final expression reduces to a convolution of the zero temperature Franck-Condon spectrum of the system, computed around the minimum energy structures of the ground and excited state PES, with a set of $N_\textrm{\scriptsize{frames}}$ $\delta$-functions placed at the vertical excitation energies of the system. Thus Eqn. \ref{eqn:vib_broadening} can be thought of as a correction to the spectrum computed via the ensemble approach, where the Gaussian broadening is replaced by a vibronic shape function accounting for the vibronic fine structure of the transition. 

In order to include explicit solvent effects at the level of the vibronic shape function, we compute the Franck-Condon spectrum for a number of representative solute-solvent conformations, where the solvent molecules are kept frozen. As mentioned previously, a rigorous way of combining the Franck-Condon and ensemble approaches would be to compute a Franck-Condon spectrum for each solute-solvent configuration sampled from the ensemble, but this protocol is beyond our current computational capabilities. Instead, we compute an average vibronic shape function from a small number of solute-solvent configurations. Denoting the frozen solvent coordinates of a given snapshot $\textbf{R}_j$ as $\textbf{R}^{\{m\}}_j$, an averaged normalized vibronic shape function $\alpha^{\textrm{\scriptsize{av}}}_{\textrm{\scriptsize{FC}}}$ is constructed via
\begin{eqnarray} \nonumber
\alpha^{\textrm{\scriptsize{av}}}_{\textrm{\scriptsize{FC}}}\left(\omega-E_{0-0}^\textrm{\scriptsize{av}};\sigma\right)=\frac{1}{c_{\textrm{\scriptsize{norm}}}} \times \\
\sum^{N_{\textrm{\scriptsize{shape}}}}_j \sigma^{\textrm{\scriptsize{vib}}}_{\textrm{\scriptsize{FC}}}\left[ T\rightarrow0 \right]\left(\omega-E_{0-0}\left[\textbf{R}^{\{m\}}_j\right];\textbf{R}^{\{m\}}_j; \sigma\right)
\label{eqn:av_shape_func}
\end{eqnarray}
where we explicitly introduce a dependence on $\textbf{R}^{\{m\}}_j$ in $\sigma^{\textrm{\scriptsize{vib}}}_{\textrm{\scriptsize{FC}}}$ and $c_{\textrm{\scriptsize{norm}}}$ is a normalization constant chosen such that $\alpha^{\textrm{\scriptsize{av}}}_{\textrm{\scriptsize{FC}}}$ integrates to 1. $E_{0-0}\left[\textbf{R}^{\{m\}}_j\right]$ is the total energy of the 0-0 transition for the Franck-Condon spectrum calculated in frozen solvent environment $\textbf{R}^{\{m\}}_j$ and $E_{0-0}^\textrm{\scriptsize{av}}$ is the average 0-0 transition energy for the $N_{\textrm{\scriptsize{shape}}}$ number of frozen solvent snapshots. Thus the spectra are averaged by assuming that their $E_{0-0}$ transition energies coincide. We justify this approach through the assumption that variations in the $E_{0-0}$ transition energy for different solvent environments are treated by the conformal sampling of vertical transition energies of the system. The resulting single, average, normalized vibronic shape function is then used in Eqn. \ref{eqn:vib_broadening}, such that the final expression for the absorption spectrum reduces to 
\begin{eqnarray} \nonumber
\sigma^{\textrm{\scriptsize{vib}}}_{\textrm{\scriptsize{MD}}}(\omega;T)= \frac{2\pi^2}{c}\frac{1}{N_{\textrm{\scriptsize{frames}}}}  \times \\
\sum^{N_\textrm{\scriptsize{frames}}}_j f_{if}\left(\textbf{R}_j\right)\alpha^{\textrm{\scriptsize{av}}}_{\textrm{\scriptsize{FC}}}\left(\omega-\omega_{if}(\textbf{R}_j)+\omega^{\textrm{\scriptsize{av}}}_{if}; \sigma\right)
\label{eqn:vib_broadening_solvent}
\end{eqnarray}
where $\omega^{\textrm{\scriptsize{av}}}_{if}=\frac{1}{N_{\textrm{\scriptsize{frames}}}} \sum^{N_\textrm{\scriptsize{frames}}}_j \omega_{if}(\textbf{R}_j)$. 

The approximate way of accounting for vibronic effects in absorption spectra of solvated dyes can thus be seen as a quantum correction to the bare vertical transitions sampled from MD by `dressing' each bare transition by an average  shape function accounting for the vibronic fine structure of the system. Since the average vibronic shape function is constructed from only a small number of representative solute-solvent conformations, such that $N_{\textrm{\scriptsize{shape}}}<<N_\textrm{\scriptsize{frames}}$, the computational cost of the combined approach is similar to directly evaluating Eqn. \ref{eqn:vert_MD}.

There are a number of situations where the approximate treatment of Eqn. \ref{eqn:vib_broadening_solvent} must necessarily break down. First, the Franck-Condon principle is only valid if the energy minima of the ground and excited PES are sufficiently close in configuration space to guarantee a sufficient overlap of ground state and excited state nuclear wave functions, which is likely not the case for very flexible dyes or systems undergoing photoisomerization. Secondly, only the quantum nature of the ground state vibrational mode is considered in the system, which is clearly invalid if the initial population of higher energy vibrational modes at the simulated temperature introduces significant changes in the structure of the Franck-Condon shape function. Lastly, representing the vibronic fine structure of the solvated system by a single average vibronic shape function is most likely not valid in systems with strong solute-solvent interactions, as the vibrational modes of solvent molecules strongly bound to the solute will likely start contributing to the vibronic shape function. It is therefore vital to compare the performance of calculating absorption spectra according to Eqn. \ref{eqn:vib_broadening_solvent} with the spectra generated from bare vertical excitations (Eqn. \ref{eqn:vert_MD}), as well as the approach of Eqn. \ref{eqn:abs_fc_solv} obtained through a rigorous separation of solute and solvent degrees of freedom in order to assess the validity of the approach in a number of different systems with varying strengths of solute-solvent interaction.  

\section{Computational details}
The aim of the present study is to reproduce the shape of experimental optical absorption spectra of the GFP anion in water,  the neutral GFP chromophore in methanol (the chromophore is protonated outside of the conjugated bridge, and is denoted neutral+), as well as Nile Red in cyclohexane, benzene, and acetone. See Fig. \ref{fig:structures_dyes} for the optimized ground state structures of all dyes studied. 

The combined ensemble plus vibronic broadening approach described in this work for modeling absorption spectra of dyes relies on three main ingredients. The first is the generation of a representative number of conformations of the dye in the chosen solvent environment for a given temperature. The second is the calculation of vertical excitation energies and oscillator strengths for the solute-solvent conformations. For these excited state calculations, the first solvation shell is explicitly included in the QM region in order to capture solvent polarization effects from first principles. Lastly, a vibronic shape function has to be computed. In this section we describe the computational details for each of the three steps involved in the calculation of the optical spectra presented here.

\subsection{MD calculations}
The ensemble solute-solvent conformations used to compute vertical excitation energies in this work are extracted from classical MD trajectories obtained using AMBER\cite{AMBER}. Although the use of a classical MD force field is a potential source of error since it does not allow for certain forms of solute-solvent interactions like proton-transfers, it enables us to access long time scales that are vital for generating a large number of uncorrelated solute-solvent conformations. In this work, we use the generalized AMBER force field (GAFF)\cite{GAFF} for all solvents apart from water, which is described by the TIP3P model\cite{TIP3P}. For the three dyes in question, we use the ANTECHAMBER program\cite{GAFF} to generate appropriate force field parameters, where the input structures are optimized in vacuum at the B3LYP/6-31+G*\cite{B3LYP} level. However, two dihedral angles of the GFP chromophores and one dihedral angle of Nile Red that are expected to couple strongly to excitation energies are reparameterized by directly fitting to the DFT PES (see Section I of the supporting information for further details). 

\begin{table}
\centering
\begin{tabular}{|c|c|c|}
\hline
System & $N^{\textrm{\scriptsize{Atoms}}}_{\textrm{\scriptsize{MD}}}$ &  $N^{\textrm{\scriptsize{Atoms}}}_{\textrm{\scriptsize{TDDFT}}}$ \\ \hline
Nile Red acetone & 13332 & 576\\
Nile Red benzene & 17970 & 561\\
Nile Red cyclohex. & 17736 & 626 \\
GFP anion water & 5340 &  570 \\
GFP neutral+ methanol & 4768 & 587 \\ \hline
\end{tabular}
\caption{Number of atoms in the MD box, as well as average number of atoms in the QM region used for the calculation of vertical excitation energies for all solvated systems studied.}
\label{tab:num_atoms}
\end{table}

The dyes are placed in large solvent boxes (see Table \ref{tab:num_atoms} for the number of atoms in the MD box for each system) and a two-step equilibration is carried out. First a 20 ps temperature equilibration in the NVT ensemble is performed to raise the temperature of the system from 0 K to 300 K. This is followed by a 400 ps volume equilibration in the NPT ensemble. Since we are interested in generating uncorrelated snapshots rather than accurate short time-scale dynamics, we run all production calculations in the NVT ensemble to guarantee a constant temperature. For each solvated system, we generate a production trajectory of 8 ns in length. Solute-solvent snapshots are extracted every 4 ps, producing a total of 2000 uncorrelated snapshots. All MD calculations are performed using a 2 fs time-step and a Langevin thermostat\cite{Langevin} with a collision frequency of 1 ps$^{-1}$. 

In order to compute the solvent-induced temperature broadening of the Franck-Condon spectrum as described by Santoro \emph{et al.}\cite{Santoro_vibronic}(see Eqn. \ref{eqn:abs_fc_solv}), which assumes a complete decoupling of solute and solvent degrees of freedom, we also perform MD simulations on Nile Red in all three solvents in which the solute is frozen in its CAM-B3LYP\cite{CAMB3LYP}/6-31G optimized ground state structure. We follow the same computational procedure as described above, with the difference that heavy positional restraints with force constants of 500 kcal\,mol$^{-1}$\,\AA$^{-2}$ are placed on all atoms belonging to the dye. For each solvent, a 2 ns production trajectory calculation is carried out and 500 snapshots of solute-solvent conformations are extracted for the computation of vertical excitations.

\subsection{Vertical excitation energies}
Vertical excitation energies are computed for all snapshots extracted from the AMBER MD simulations using TDDFT. For all systems studied it is found to be sufficient to converge the lowest three singlet excitations in order to fully cover all excitations in the visible region of the spectrum. 

Solvent polarization effects and the solute-solvent interaction on the excitation energies are modeled from first principles by explicitly including a significant part of the solvent environment in the QM region for the TDDFT calculation. The size of the QM region is controlled by a single cutoff radius $R_{\textrm{cut}}$=8\,\AA\, for all solvated systems, where all solvent molecules with a center of mass that is within $R_{\textrm{cut}}$ of any atom of the dye are treated fully quantum mechanically. An example QM region for a snapshot of Nile Red in acetone can be found in Fig. \ref{subfig:qm_region} and Table \ref{tab:num_atoms} details the average number of QM atoms included in the TDDFT calculations for each solvated system. All solvent molecules with centers of mass lying outside of the cutoff radius are included into the TDDFT calculation in the form of electrostatic point charges, where the charges of the solvent atoms are fixed to those specified by the AMBER force field. We have performed a number of test calculations on all solvated systems where the cutoff radius is increased from 8\,\AA\, to 10\,\AA\, in order to confirm that the vertical excitations energies are converged with respect to the size of the QM region (see Section IIB of the supporting information). 

All TDDFT calculations of vertical excitation energies are carried out using the TeraChem software package\cite{terachem1,terachem2}. The Tamm-Dancoff approximation\cite{TDA,TDA_head_gordon} is used throughout and all calculations are performed at the CAM-B3LYP/6-31G\cite{CAMB3LYP} level of theory. The basis set is too small to converge the absolute value of excitation energies in the chosen systems, which is especially apparent for the GFP anion as previously noted in the literature\cite{Krylov_paper}. However, we show in a number of calculations on a smaller subset of snapshots that although larger 6-31G* and 6-31+G* basis sets cause a significant red-shift of excitation energies, this shift is relatively systematic, such that the computed shape and width of the spectrum of vertical absorption energies is largely unchanged (see Section IIA of the supporting information). Since the focus of this study is the modeling of the shape of absorption spectra rather than the absolute position of absorption maxima, and given the large number of individual TDDFT calculations that need to be performed for this work, the smaller 6-31G basis set is considered to be sufficient and is used throughout. 

It is worth pointing out that standard approaches accounting for temperature-dependent and solvent-induced broadening in a phenomenological way (see Eqn. \ref{eqn:pcm_spectrum} and \ref{eqn:abs_fc_solv}) generally assume the broadening to be Gaussian in nature, such that it can be described by a single parameter $\sigma(T)^2$. A broadening purely derived from the MD sampling of vertical excitation energies (see Eqn. \ref{eqn:vert_MD} and \ref{eqn:vib_broadening_solvent}), however, does not make any assumption about the shape of the distribution of vertical excitation energies. To assess how closely the excitation energies sampled with MD follow a Gaussian distribution, we compute the generalized third and fourth moments for all systems. The third moment, referred to as the skewness, is used as a measure of the asymmetry of the distribution, whereas the generalized fourth moment, the kurtosis, measures the heaviness of the tails. Both measures taken together reveal how non-Gaussian the distribution of vertical excitation energies is and allow us to assess the validity of assuming a phenomenological Gaussian broadening determined by a single parameter $\sigma(T)^2$ (see Section IV of the supporting information for further details).  

\subsection{Vibronic shape functions}
\label{sec:vibronic_computation}
For all systems studied, the TDDFT S$_1$ excited state is the only state with significant oscillator strength and is thus the only state treated with the vibronic shape function approach (Eqn. \ref{eqn:vib_broadening_solvent}), whereas a Gaussian line shape is assumed for all other excitations (Eqn. \ref{eqn:vert_MD}). 

All Franck-Condon spectra for Nile Red and the GFP neutral+ chromophore are computed using Gaussian09\cite{Gaussian} at the CAM-B3LYP/6-31G level of theory in order to provide a consistent treatment with the computation of vertical absorption energies. Although the spectra for all dyes show some dependence on the basis set used (see Section IIIA of the supporting information), we note that the smaller 6-31G basis set correctly captures the main features of the spectra for Nile Red and the GFP neutral+ chromophores. For the GFP anion chromophore, the changes in the vibronic spectra with respect to basis set choice are somewhat more pronounced, and mainly involve a shift of spectral weight from the dominant vibronic peak to higher energy structures. For this reason, the larger 6-31+G* basis set is used to compute the Franck-Condon spectra for the GFP anion. All Franck-Condon spectra are broadened by a Gaussian function with standard deviation $\sigma=0.0105$ eV. We find that this broadening parameter, together with a sampling of vertical excitation energies for 2000 snapshots, is sufficient to yield smooth absorption spectra for all systems studied in this work, while preserving all of the dominant features in the vibronic fine structure of the Franck-Condon spectrum.  

The choice of CAM-B3LYP as a functional is made to be fully consistent with the computation of vertical excitation energies. However, we note that the vibronic spectra of Nile Red are rather sensitive to the choice of exchange-correlation functional, whereas the GFP chromophores show considerably less variation (see Section IIIB of the supporting information for a comparison between the two range-separated hybrid functionals CAM-B3LYP and $\omega$B97X\cite{omegaB97X} for all three chromophores). For Nile Red, comparisons of different TDDFT functionals to higher order quantum chemistry approaches\cite{nile_red_higher_order} suggest the S$_1$ state is very sensitive to the treatment of long-range exchange. Furthermore, previous calculations of Franck-Condon spectra using TDDFT\cite{Mennucci_nile_red} suggest that the CAM-B3LYP functional performs well in comparison to experimental results. A similar study has been carried out for the GFP anion in vacuum, where it was found that the CAM-B3LYP functional produces Franck-Condon spectra in close agreement to those computed with CASPT2 and XMCQDPT2\cite{Santoro_GFP_gas_phase}. We therefore use the CAM-B3LYP functional for all Franck-Condon spectra of both the GFP chromophores and Nile Red. However, we notice that the sensitivity of Franck-Condon spectra to the shape of the ground state and especially the S$_1$ PES means that, wherever possible, the choice of functional should be informed by comparisons to more accurate $\emph{ab-initio}$ methods. It has been demonstrated recently that the vibrational reorganization energy is a good criterion for assessing the accuracy of Franck-Condon spectra and thus allows for an efficient comparison between $\emph{ab-initio}$ and DFT methods\cite{vibronic_spectra_functional}.

The vibronic broadening approach of Eqn. \ref{eqn:vib_broadening_solvent} assumes that only the ground state vibrational mode of the electronic ground state is initially occupied and contributes to the Franck-Condon shape function; therefore all temperature effects are treated classically from the MD. We test this assumption by including temperature effects on the Franck-Condon spectra using the FCclasses code\cite{FCclasses1,FCclasses2,FCclasses3}, which populates a set of initial states with Boltzmann weights. We confirm that the contribution of higher vibrational states of the ground state PES is small at T=300 K for all dyes studied in this work and only slightly broadens the spectrum rather than leading to substantial changes in the vibronic fine structure (see Section IIIC of the supporting information).

\begin{figure}
\centering
\subfloat[Vertical excitation energies \label{subfig:qm_region}]{\resizebox{0.25\textwidth}{!}{\includegraphics{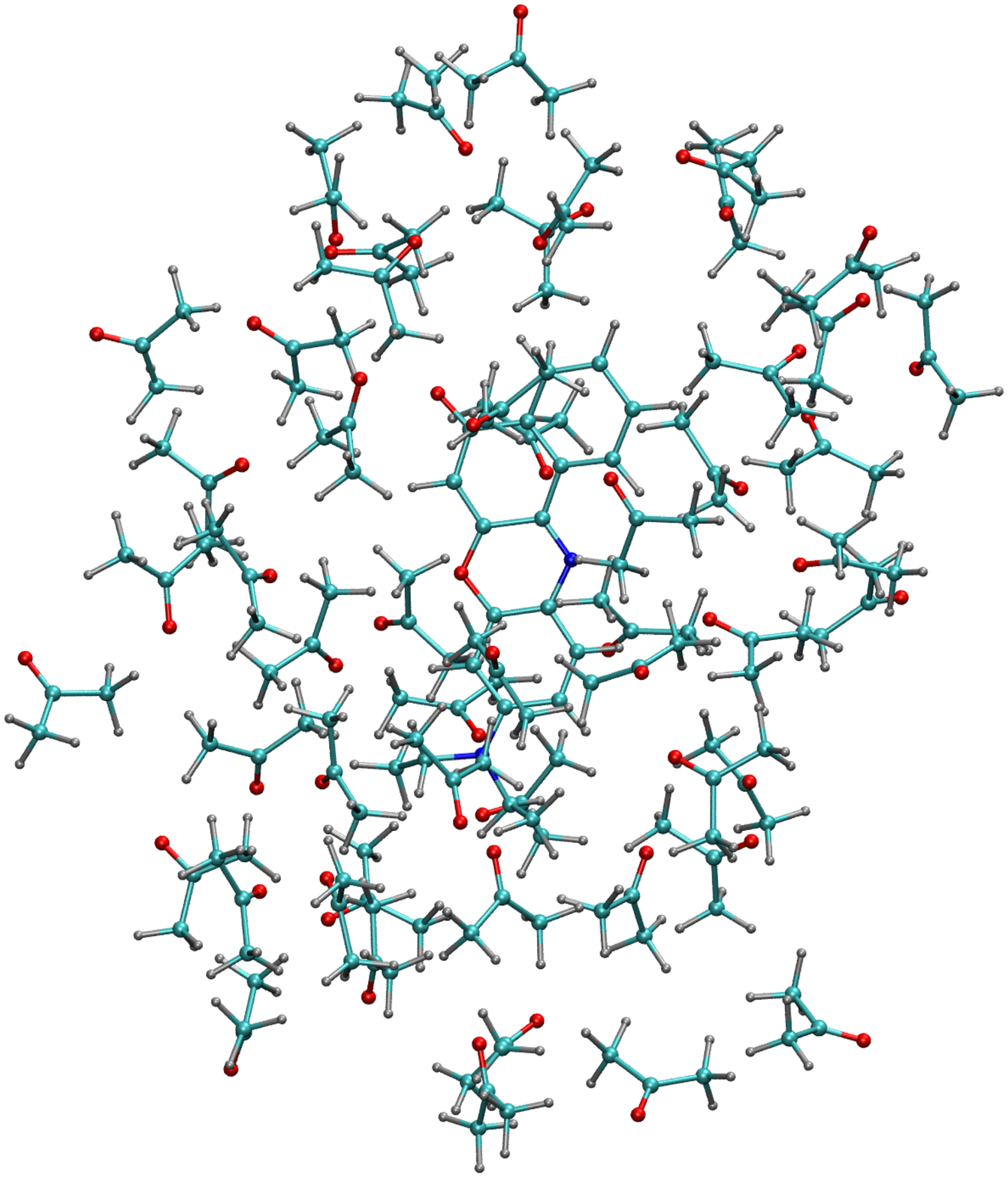}}}
\subfloat[Vibronic shape function \label{subfig:qm_region_vibronic}]{\resizebox{0.25\textwidth}{!}
{\includegraphics{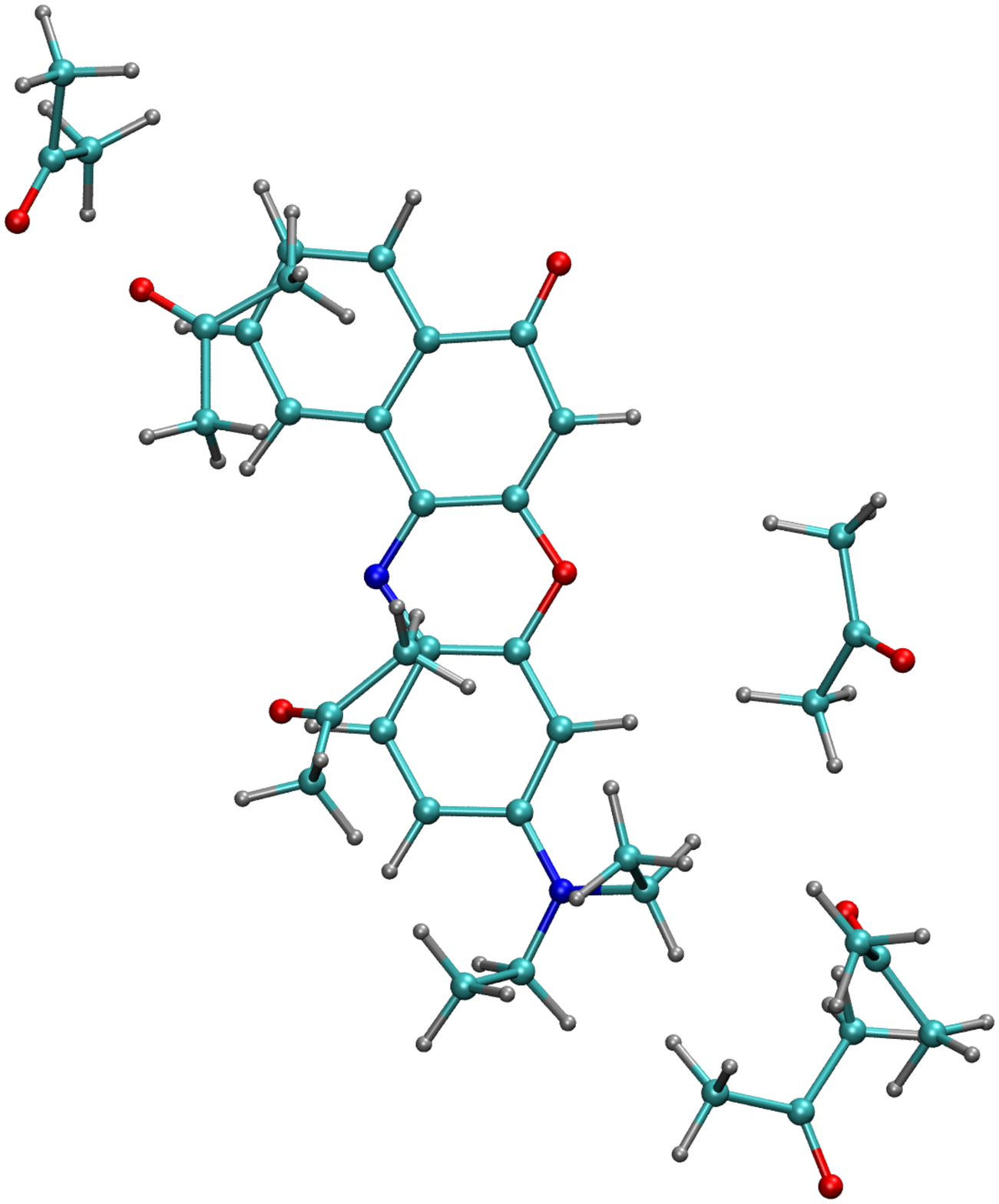}}}
\caption{The two different QM regions used in this work, as illustrated on example snapshots of Nile Red in acetone. For the computation of vertical excitation energies, the QM region includes the entire first solvation shell, while for the computation of the vibronic shape function in a frozen solvent environment, only a small number of the closest solvent molecules are retained in the QM region.}
\end{figure}

\begin{figure}
\centering
\subfloat[GFP anion in water \label{subfig:shape_func_anion_water}]{\resizebox{0.4\textwidth}{!}{\includegraphics[angle=270]{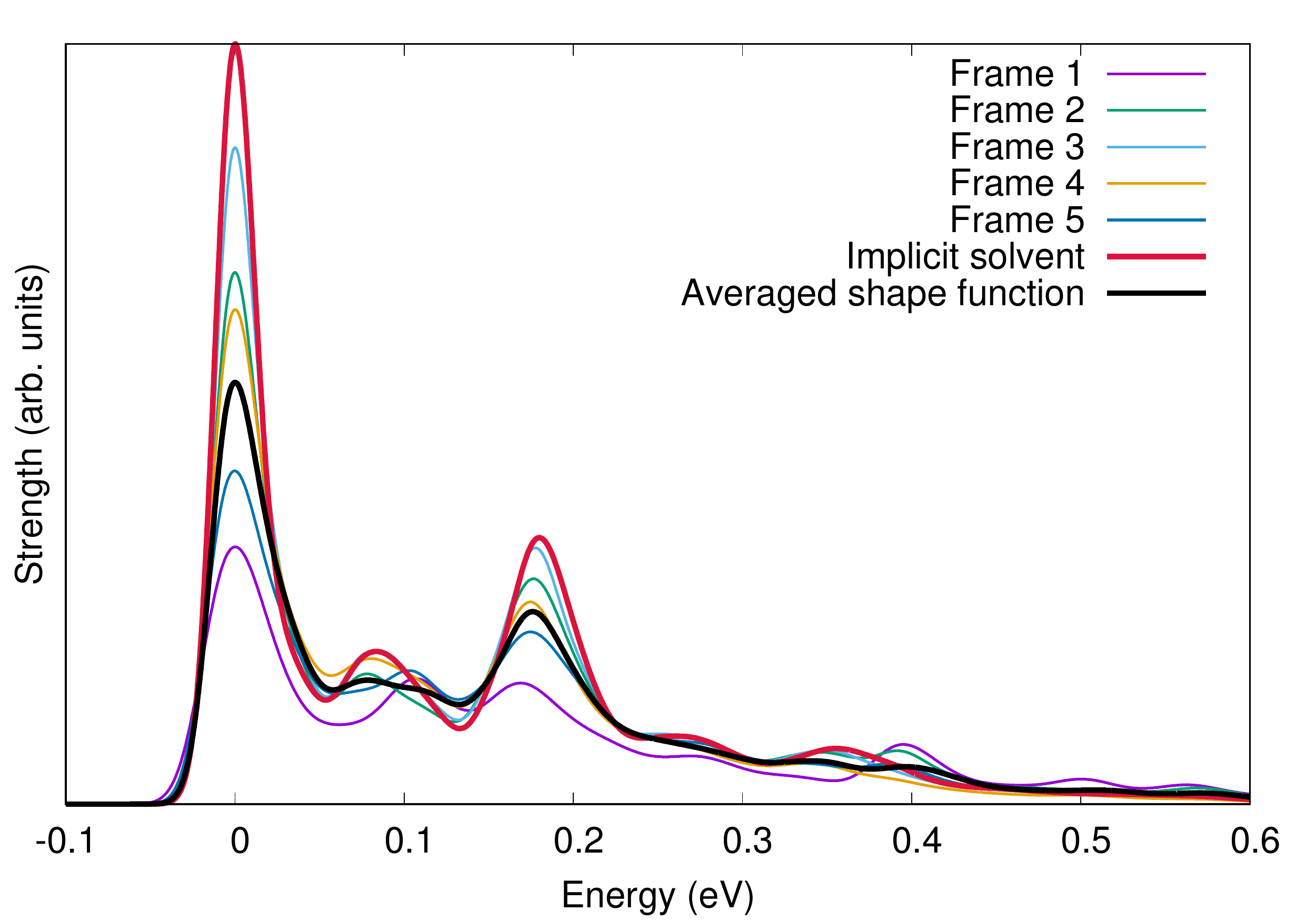}}} \\
\subfloat[Nile Red in acetone \label{subfig:shape_func_nile_red_acetone}]{\resizebox{0.4\textwidth}{!}
{\includegraphics[angle=270]{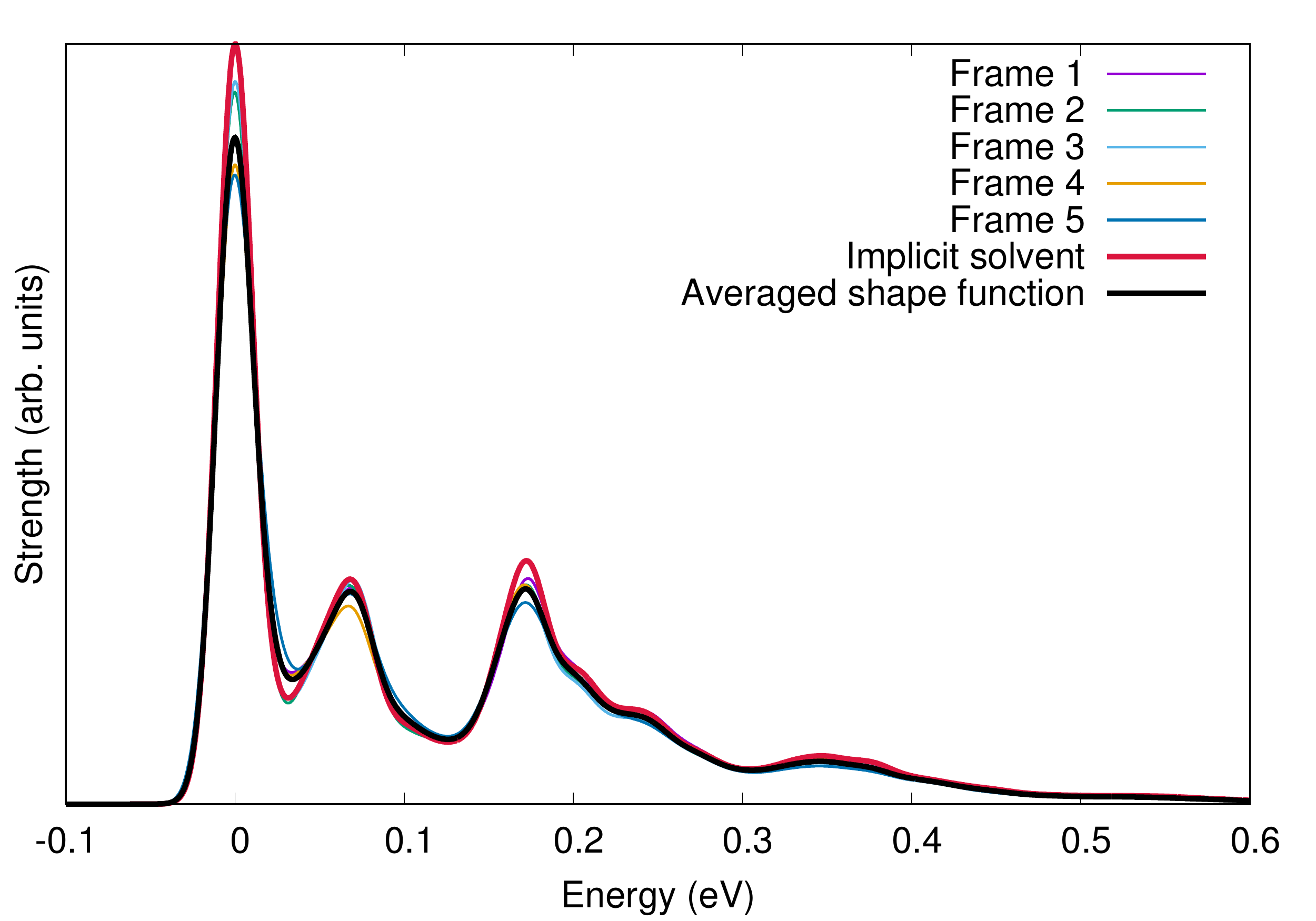}}}
\caption{Vibronic shape functions for the GFP anion in water and Nile Red in acetone, as calculated in implicit solvent as well as in frozen solvent environments taken from five uncorrelated MD snapshots. The vibronic shape function used to construct the absorption spectrum is the average of the vibronic spectra from the five individual frozen solvent snapshots and is also plotted for comparison. All spectra are shifted such that the 0-0 transition is at 0 eV. A Gaussian broadening of $\sigma=0.0105$ eV is applied to all Franck-Condon spectra. }
\label{fig:shape_func_acetone}
\end{figure}

The vibronic shape function is computed according to Eqn. \ref{eqn:av_shape_func} as an average of five explicit solute-solvent conformations chosen from the 2000 uncorrelated MD snapshots. The high computational cost associated with the ground and excited state geometry optimizations necessary for computing Franck-Condon spectra makes it prohibitive to use the same QM region in our calculations that is used in the calculation of vertical excitation energies. Rather, we limit ourselves to a few solvent molecules closest to the dye as defined by the distance between the center of mass of the solvent and any atom of the dye, such that the total QM region size for each snapshot studied ranges from $\approx$ 80-130 atoms depending on the system in question (see Fig. \ref{subfig:qm_region_vibronic} for an example QM region used in the computation of the vibronic shape function of Nile Red in acetone).  The explicit solvent representation is expected to include the solvent molecules that interact most strongly with the solute, such as hydrogen-bonded solvent molecules in the case of the GFP anion in water, but is not large enough to provide a full first solvation shell. For this reason, the QM region is surrounded by a PCM in order to represent the long-range electrostatic effects on the solute. Geometry optimizations of the dye in the ground and excited state keep the QM solvent atoms frozen and dispersion interactions between the molecules are accounted for using Grimme's D3 empirical dispersion correction\cite{empirical_dispersion}. 

The average vibronic shape function for both Nile Red in acetone and the GFP anion in water, along with the Franck-Condon spectra of the five individual frozen solvent conformations, as well as the Franck-Condon spectrum calculated in implicit solvent, can be found in Fig. \ref{fig:shape_func_acetone}. For Nile Red in acetone, the different frozen solvent environments introduce very minor changes to the computed Franck-Condon spectra, such that the average shape function closely resembles the shape function computed in implicit solvent. For the GFP anion in water, however, the different solvent environments introduce considerable changes to the Franck-Condon spectra, most notably a redistribution of spectral weight from the 0-0 transition to higher energy vibronic states for a number of snapshots. The results can be straightforwardly interpreted by considering the strength of solute-solvent interactions for the two systems. As expected, the aprotic solvent acetone interacts weakly with Nile Red, yielding explicit solvent Franck-Condon spectra in close agreement with the spectrum computed in PCM. For the GFP anion in water, solute-solvent interactions are considerably stronger and the Franck-Condon spectrum computed in PCM does not provide a good approximation to the spectrum computed in different explicit solvent environments. 

\section{Results and discussion}
Throughout this work, we compare spectra computed using the ensemble approach with Gaussian broadening (Eqn. \ref{eqn:vert_MD}) with our combined ensemble plus Franck-Condon broadening (Eqn. \ref{eqn:vib_broadening_solvent}). We first compute the vertical excitation energies for the uncorrelated MD snapshots as prescribed in the previous section. The vertical absorption spectrum is then constructed in the standard ensemble approach by applying a Gaussian broadening of $\sigma=0.0105$ eV to all calculated transitions. For the combined approach, we instead take the same vertical transition energies and convolute the bright S$_1$ state with the average Franck-Condon vibronic shape function evaluated in the zero-temperature limit, where a Gaussian broadening of $\sigma=0.0105$ eV is applied to all individual vibronic transitions. For the other vertical excitations, which are considerably less bright than the S$_1$ state, a Gaussian lineshape with identical standard deviation is assumed. Both absorption spectra still account for temperature effects in a purely classical way through the conformational sampling of the chromophore, however, the combined approach also takes into account the quantum nature of the nuclei through the vibronic shape function. 

The focus of this work is the reproduction of the experimental spectral shape, as well as thermally induced and solvent broadening effects without resorting to any phenomenological broadening parameters fitted to experiment. We do not focus on reproducing the absolute position of absorption maxima, which are very sensitive to errors in the chosen DFT functional and basis set size\cite{Krylov_paper}. For this reason, all simulated spectra are shifted and scaled in order to align with the experimental absorption maxima. 

\subsection{GFP chromophores in solution}
We first focus on the two GFP model chromophores. The anion is solvated in water and the neutral+ is solvated in methanol, so we expect strong hydrogen bonding between the solute and solvent. The GFP chromophore in aqueous environment is known to change its protonation state depending on the pH of the solution\cite{GFP_protonation_water,GFP_protonation_ethanol}, with a pH=7 yielding a mixture of the neutral and the anionic form\cite{GFP_protonation_water,GFP_anion_experiment}. Since the classical force field approach used in this work cannot capture protonation and deprotonation events, the GFP anion spectrum is simulated in pure water, but is compared to experimental results obtained in aqueous solutions of pH=13\cite{GFP_anion_experiment}, which corresponds to the GFP chromophore being completely in its anionic form. Given that the MD are run in pure water, we ignore any potential effects of counter ions on the computed spectrum. 

\begin{figure}
\centering
\subfloat[GFP anion in water \label{subfig:gfp_anion_water}]{\resizebox{0.45\textwidth}{!}{\includegraphics[angle=270]{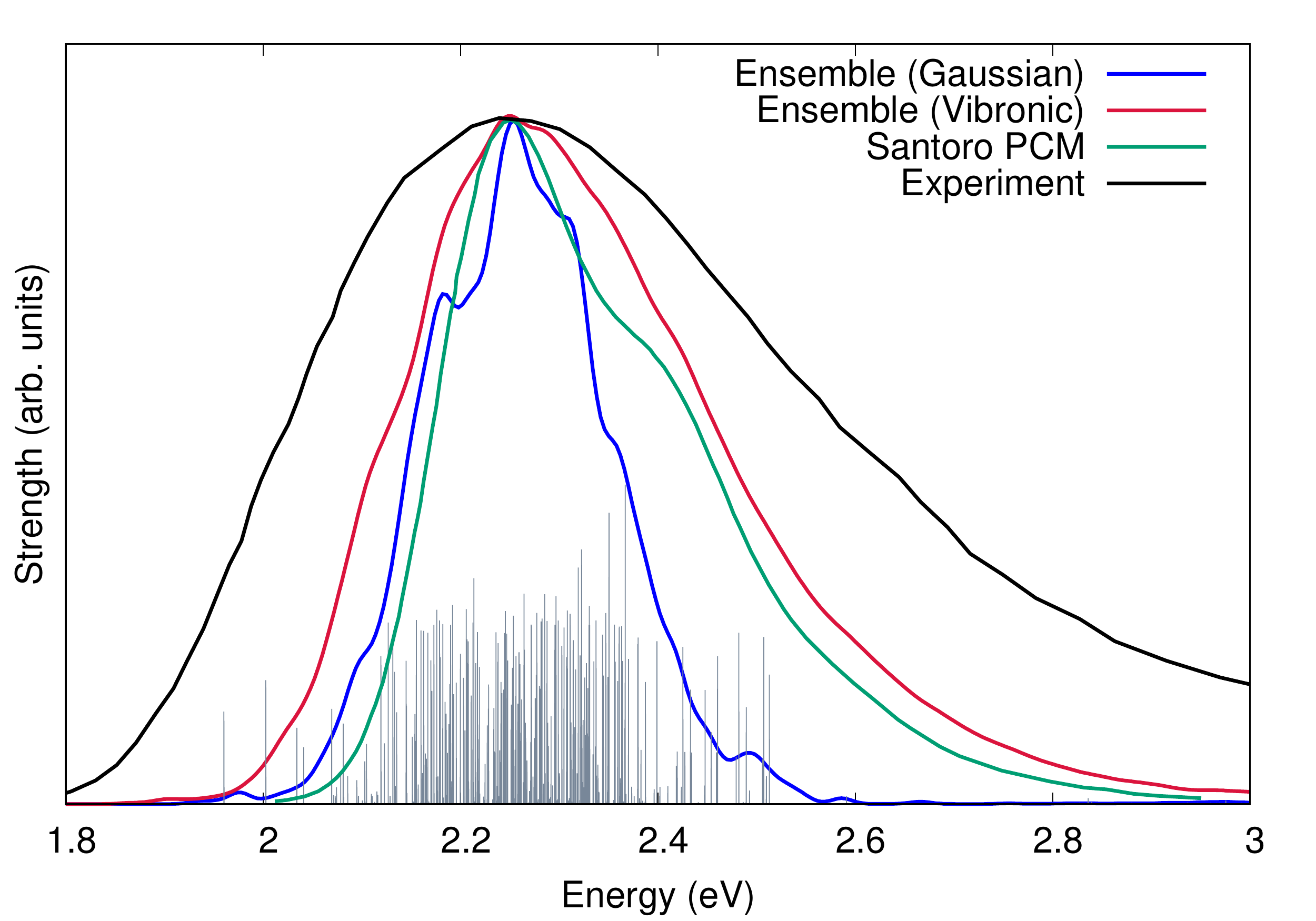}}} \\
\subfloat[GFP neutral+ in methanol\label{subfig:gfp_neutral_plus_methanol}]{\resizebox{0.45\textwidth}{!}
{\includegraphics[angle=270]{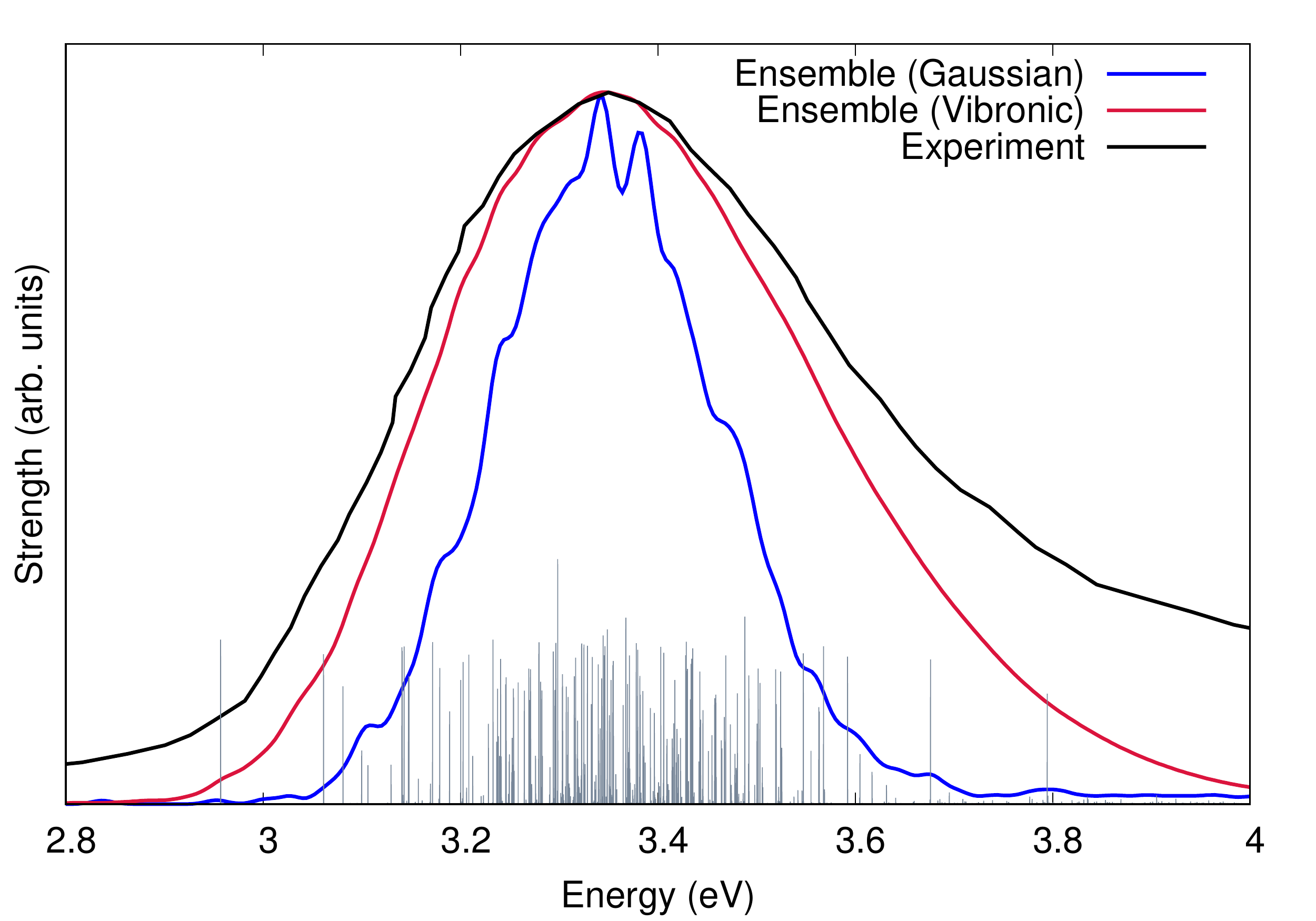}}}
\caption{Absorption spectra for the GFP anion in water and GFP neutral+ in methanol, as calculated with both an ensemble approach with Gaussian broadening and the combined ensemble plus vibronic broadening approach, and comparison with experiment\cite{GFP_anion_experiment,GFP_neutral_plus_experiment}. The spectrum obtained by Santoro \emph{et al}.\cite{Santoro_GFP_solution} for the GFP anion using a temperature-dependent Franck-Condon spectrum broadened by a Gaussian with a standard deviation derived from PCM calculations is also displayed. The experimental spectrum for the GFP anion is taken in an aqueous solution of NaOH$_{(aq)}$ with pH=13. The spectra are computed from vertical excitation energies of 2000 uncorrelated MD snapshots and a broadening of $\sigma=0.0105$ eV is used in both broadening approaches. The bare vertical excitation energies obtained from the MD sampling are shown as a stick spectrum. Calculated spectra are shifted and scaled in order to match experimental data.}
\label{fig:gfp_solvent}
\end{figure}

We first compute the bare vertical excitation energies from the 2000 solute-solvent conformations extracted from MD trajectories for each system, which are shown in Fig. \ref{fig:gfp_solvent} as vertical lines with the height of the line representing the oscillator strength of the transition. The MD sampling accounts for temperature effects in solute and solvent conformations, as well as explicit solute-solvent interactions such as hydrogen bonding, whereas the large QM region used in the TDDFT calculation captures solvent polarization effects on the electronic excitations. However, when comparing the absorption spectra produced by the ensemble approach to the experimental results (see Fig. \ref{fig:gfp_solvent}), we find that the simulated spectra are considerably too narrow and do not reproduce the correct asymmetry. An analysis of the vertical excitation energies (see Section IV of the supporting information) reveals that although for both solutes the distributions are non-Gaussian and show a tendency to from tails in the higher energy, the observed skewness is not sufficient to yield the asymmetric experimental spectrum of the GFP anion in water. 

We construct effective vibronic shape functions by averaging over several Franck-Condon spectra computed in frozen solvent environments as described in Section \ref{sec:vibronic_computation}. Although freezing the solvent molecules does not allow for any coupling of vibrational modes of the solvent to the electronic excitation of the solute, it provides an approximate description of the influence of the explicit solvent environment on vibrational modes of the solute. The vibronic shape functions of both chromophores computed for five different frozen solvent environments, the average shape function, and the Franck-Condon spectra constructed in implicit solvent can be found in Fig. \ref{subfig:shape_func_anion_water} for the GFP anion in water and in Section IIID in the supporting information for the neutral+ chromophore in methanol. In contrast to Nile Red in acetone (see Fig. \ref{subfig:shape_func_nile_red_acetone}), the inclusion of a frozen explicit solvent environment in the computation of the Franck-Condon spectra does yield significant changes in relative peak intensities as compared to spectra computed in PCM. The larger variability between the Franck-Condon spectra of individual frozen solvent conformations can likely be ascribed to the much stronger solute-solvent interaction for the GFP chromophores in methanol and water, as both solvents form strong hydrogen bonds with the chromophores. 

Once the average vibronic shape function is constructed for both solvated systems, the absorption spectrum can be computed in the combined ensemble plus vibronic broadening approach. The results in Fig. \ref{fig:gfp_solvent} show a significant improvement over the ensemble approach for both GFP chromophores. For the GFP neutral+ variant, both the shape and width of the spectrum are in good agreement with experiment, with the main discrepancy being that the long high energy tail lacks some intensity for energies larger than $\approx$ 3.7 eV. This failure is likely due to other excited states beyond the bright S$_1$ state contributing to the spectrum that are not recovered in the correct position and with the correct relative intensity in our computation of the vertical excitation energies. 

For the GFP anion in water, the inclusion of vibronic effects also yields significant broadening and improvement of the general shape of the spectrum. However, the resulting simulated spectrum is still considerably too narrow. Although it is difficult to isolate the source of this failure in our calculations, we note that MD sampling using classical force fields might not be fully appropriate in a system with strong hydrogen bonding since conformations corresponding to shared protons between the anion and a water molecule cannot be adequately sampled. These effects could potentially be captured by making use of ab-initio MD, as well as including nuclear quantum effects in the dynamics of the water molecules by using path integral molecular dynamics (PIMD) approaches\cite{PIMD1,PIMD2}, but these computationally more expensive MD methods are beyond the scope of the current work. 

Apart from the potential limitations of our conformational sampling using classical force fields, we note that the combined ensemble plus Franck-Condon approach does seem to yield good results even in systems with strong solute-solvent interactions. In particular, we correctly capture the overall shape of both solution-phase absorption spectra. Although the Franck-Condon spectra for both dyes show significant vibronic fine structure with secondary peaks beyond the 0-0 transition, these features wash out in the MD sampling to produce a smooth vibronic tail in solution, in agreement with experiment. 

The GFP anion results are also interesting from the perspective that the system has already been studied previously by calculating the full temperature-dependent Franck-Condon spectrum (Eqn. \ref{eqn:abs_fc_solv}) with a broadening derived from a PCM model\cite{Santoro_GFP_solution}. Computing the solvent reorganization energy from the difference in excited state energies of optimized solute geometries assuming equilibrium and non-equilibrium solvation allows one to calculate a Gaussian broadening parameter $\sigma(T)$ following Marcus theory. The results obtained in \cite{Santoro_GFP_solution} following this approach are also displayed in Fig. \ref{fig:gfp_solvent}. As can be seen, the Gaussian broadening derived from the solvent reorganization energy computed in PCM yields a spectrum that is even more narrow than our results using the ensemble plus vibronic broadening approach. Furthermore, the spectrum shows significant vibronic fine structure in form of a shoulder that is smeared out into a smooth vibronic tail in our results. The failure in obtaining the correct spectral shape and width in \cite{Santoro_GFP_solution}  is attributed to the fact that the GFP chromophore twists in solution, yielding more broadening than would be expected from a pure QM treatment of temperature effects through the population of ground state vibrational modes. The purely classical treatment of temperature fluctuations through the MD sampling of solute-solvent conformations used in this work, on the other hand, seems to produce good spectral shapes for both systems.

It is worth pointing out that the significant variability observed in the vibronic spectral shapes obtained for different frozen solvent environments in the case of the GFP anion means that the assumption of a single vibronic fine structure represented by an average shape function can only be of limited validity in this system. For this reason, we test for the stability of the computed absorption spectra with respect to the choice of vibronic shape functions (see Section V of the supporting information). We find that although Franck-Condon spectra derived from different explicit solvent environments differ significantly, these differences yield relatively minor changes in the computed absorption spectra. This can be taken as good evidence that the use of an average vibronic shape function holds approximately in the system. The fact that the simple computational approach based on a single average vibronic shape function still yields relatively good results in this case of strong solute-solvent interactions gives reason to expect the method to perform well in systems with weaker interactions. 

\subsection{Nile Red in solution}
We next focus on the Nile Red chromophore solvated in acetone, benzene, and cyclohexane. The experimental results for these solvent environments are interesting, as the optical spectra show a significant degree of variability. The experimental spectrum in the non-polar solvent cyclohexane shows a double peak structure\cite{Nile_red_experiment_spectrum} that is vibronic in nature\cite{Mennucci_nile_red}; however, this feature changes to a flattened top in benzene and disappears completely in acetone. Furthermore, all spectra show a strong asymmetry, with long tails in the high energy wavelengths. Given the aprotic nature of the three solvents, solute-solvent interactions are expected to be considerably weaker than for the GFP chromophores in methanol and water. 
\begin{figure}
\centering
\subfloat[Nile Red in cyclohexane \label{subfig:nile_red_cyclohex}]{\resizebox{0.45\textwidth}{!}{\includegraphics[angle=270]{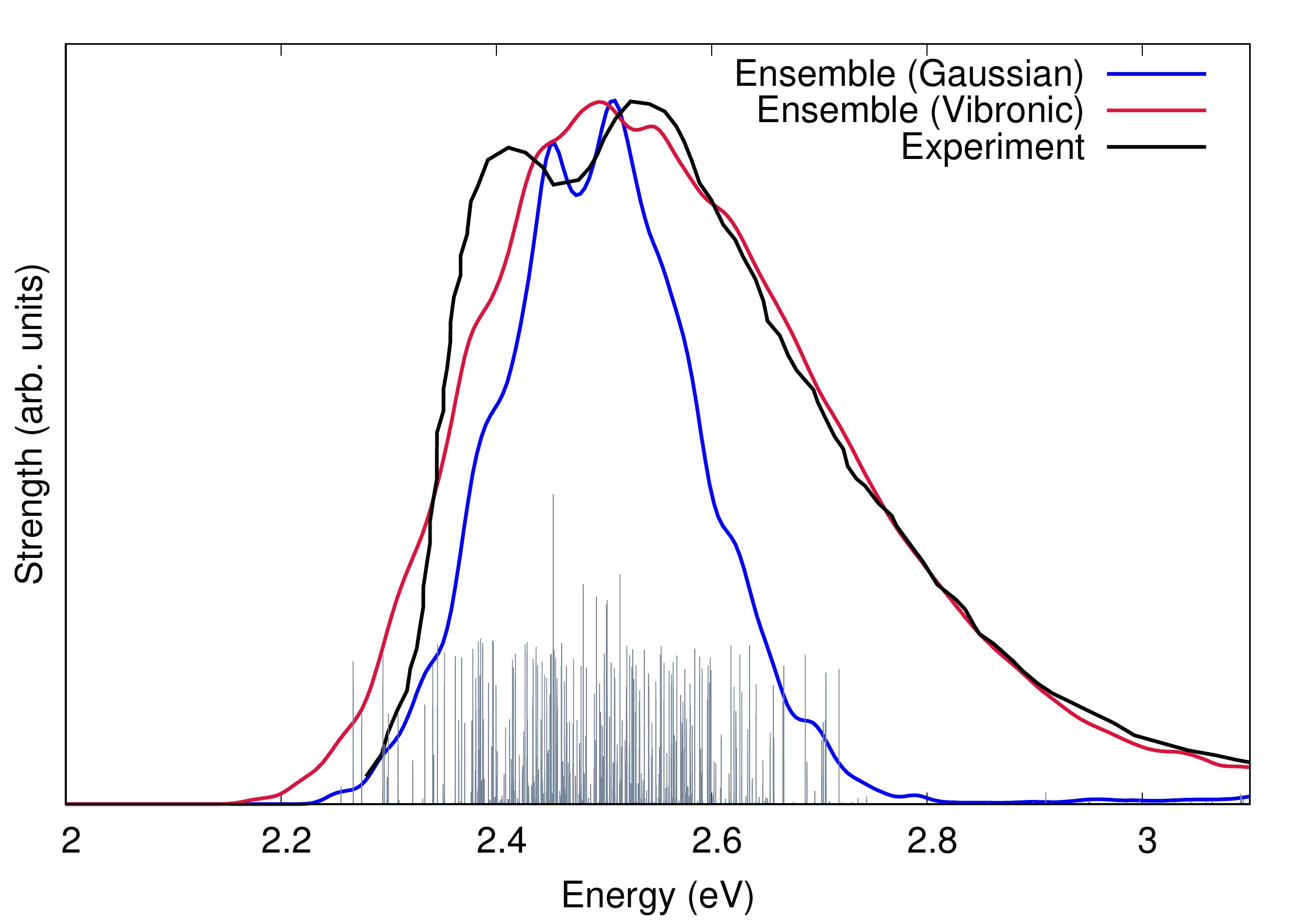}}} \\
\subfloat[Nile Red in benzene\label{subfig:nile_red_benzene}]{\resizebox{0.45\textwidth}{!}
{\includegraphics[angle=270]{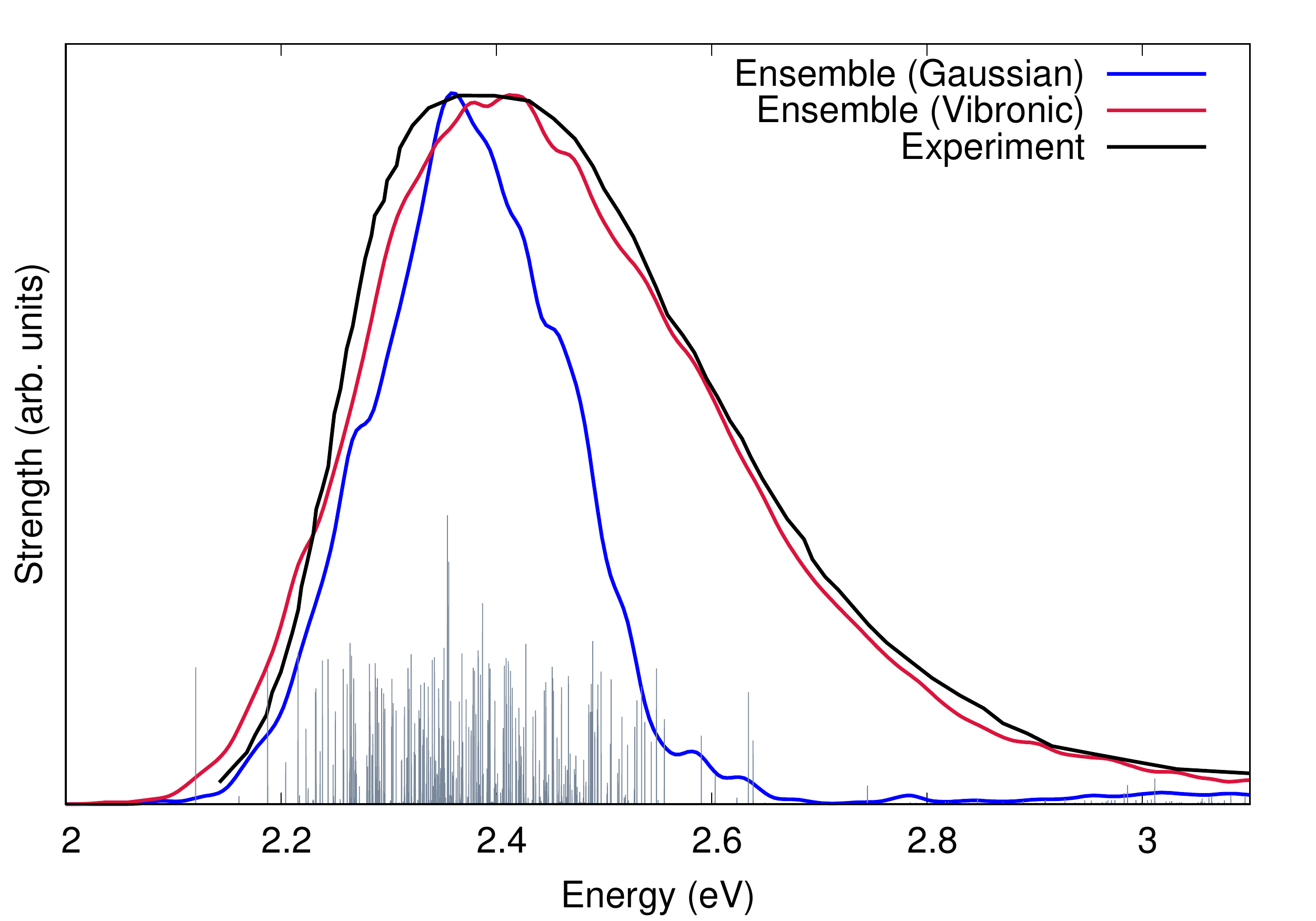}}} \\

\subfloat[Nile Red in acetone \label{subfig:nile_red_acetone}]{\resizebox{0.45\textwidth}{!}
{\includegraphics[angle=270]{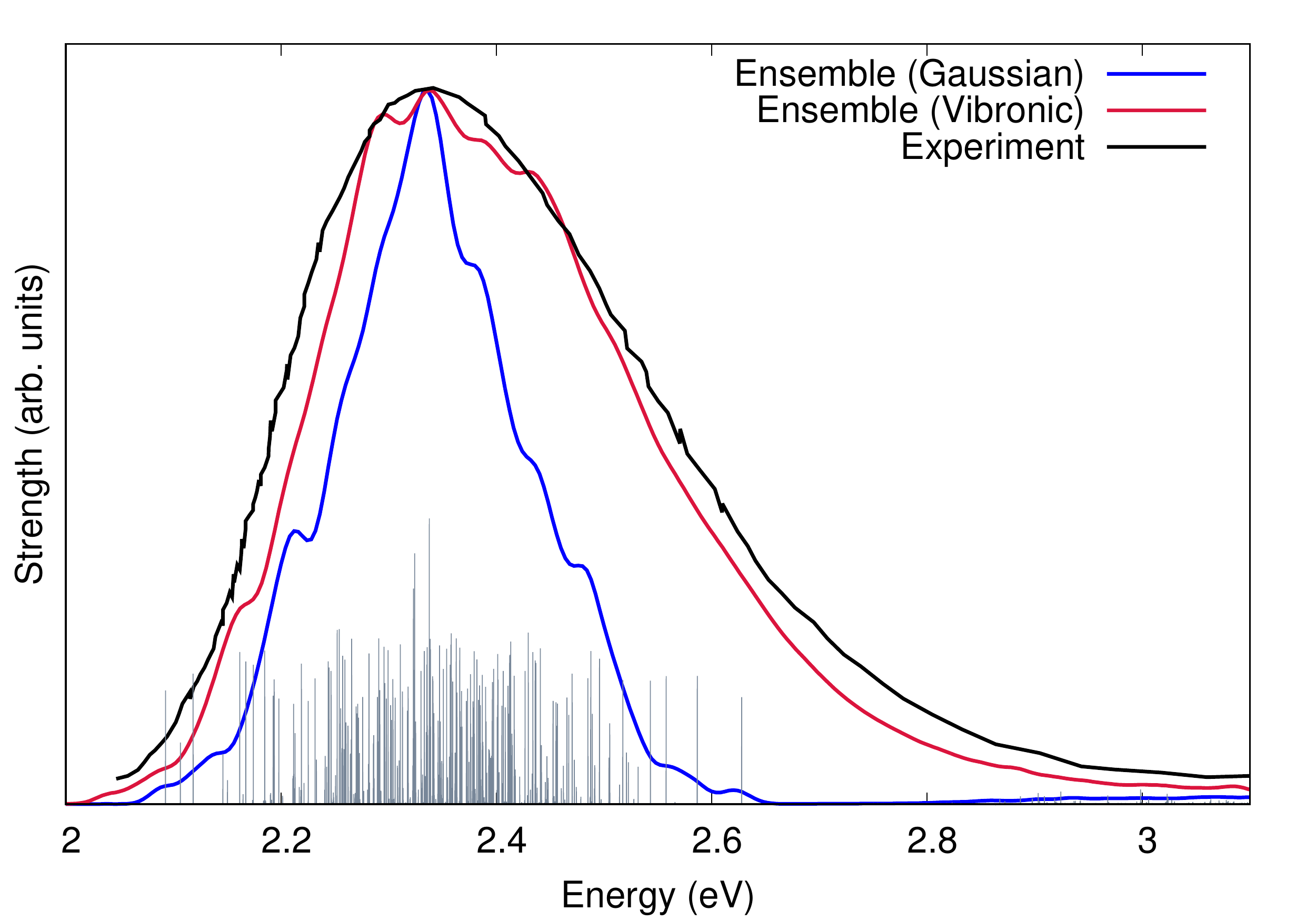}}}
\caption{Absorption spectra for Nile Red in three different solvents, as calculated with both an ensemble approach with Gaussian broadening and the combined ensemble plus vibronic broadening approach, and comparison with experiment\cite{Nile_red_experiment_spectrum}. The spectra are computed from vertical excitation energies of 2000 uncorrelated MD snapshots and a broadening of $\sigma=0.0105$ eV is used in both broadening approaches. The bare vertical excitation energies obtained from the MD sampling are shown as a stick spectrum. Calculated spectra are shifted and scaled in order to match experimental data. }
\label{fig:nile_red_spectra}
\end{figure}

The spectra computed with the ensemble approach and the combined approach, along with experimental spectra\cite{Nile_red_experiment_spectrum}, are shown in Fig. \ref{fig:nile_red_spectra} for all three solvents. The vibronic shape functions can be found in Fig. \ref{subfig:shape_func_nile_red_acetone} for Nile Red in acetone and in Section IIID of the supporting information for Nile Red in cyclohexane and benzene. With the ensemble approach,  the spectra are again significantly too narrow and the strong asymmetry and the long tails in the high energy part of the spectra are completely missed. An analysis of the distribution of vertical excitation energies (see Section IV of the supporting information) reveals again that although the distributions are markedly non-Gaussian, they only show moderate positive skew which cannot account for the strong asymmetry of the experimental spectra. This analysis suggests that, just as for the GFP chromophores, the experimental spectrum cannot be reproduced by only considering the bare vertical excitation energies of solute-solvent conformations as in the ensemble approach. 

When accounting for the vibronic fine structure of the bright S$_1$ excitation by the combined ensemble plus Franck-Condon approach, the agreement with experiment improves significantly. For all solvent environments, the width of the spectrum and the long vibronic tail are captured very well by the vibronic broadening approach. The change in the fine structure of the spectrum when changing the solvent environment, however, is recovered less well. Most notably, our results for cyclohexane show no clear double peak structure, and the flattened top of the spectrum for benzene is only approximately recovered.  For Nile Red in cyclohexane, the cause for the failure to reproduce the double-peak structure can potentially be ascribed to issues with the distribution of underlying vertical excitation energies. The experimental spectrum for Nile Red in cyclohexane has a very sharp onset and shows a steep increase in intensity until the first peak at $\approx 2.4$ eV. In contrast, the computed vertical excitation energies produce a much more gradual absorption onset. Since the 0-0 transition is the most intense transition in the Franck-Condon spectrum for Nile Red (see section IIID in the supporting information), the slow onset of the absorption spectrum from vertical excitation energies alone is directly carried over to the spectrum including vibronic effects. A second potential error is that, unlike for Nile Red in the other two solvents, the Franck-Condon spectra for Nile Red in cyclohexane show some degree of variability (see section IIID of the supporting information). In order to test how our  results depend on the vibronic shape function, we examined the computed spectrum of Nile Red in cyclohexane for different shape functions (see section V of the supporting information). The spectrum is found to be very robust under the observed variability of the vibronic shape functions for different solute-solvent environments. This finding points to the failure in reproducing the experimental double peak structure being a failure of the underlying force field in producing an accurate set of solute-solvent conformations for the computed vertical excitation energies. 

\begin{figure}
\centering
\subfloat[Nile Red in cyclohexane \label{subfig:nile_red_cyclohex2}]{\resizebox{0.45\textwidth}{!}{\includegraphics[angle=270]{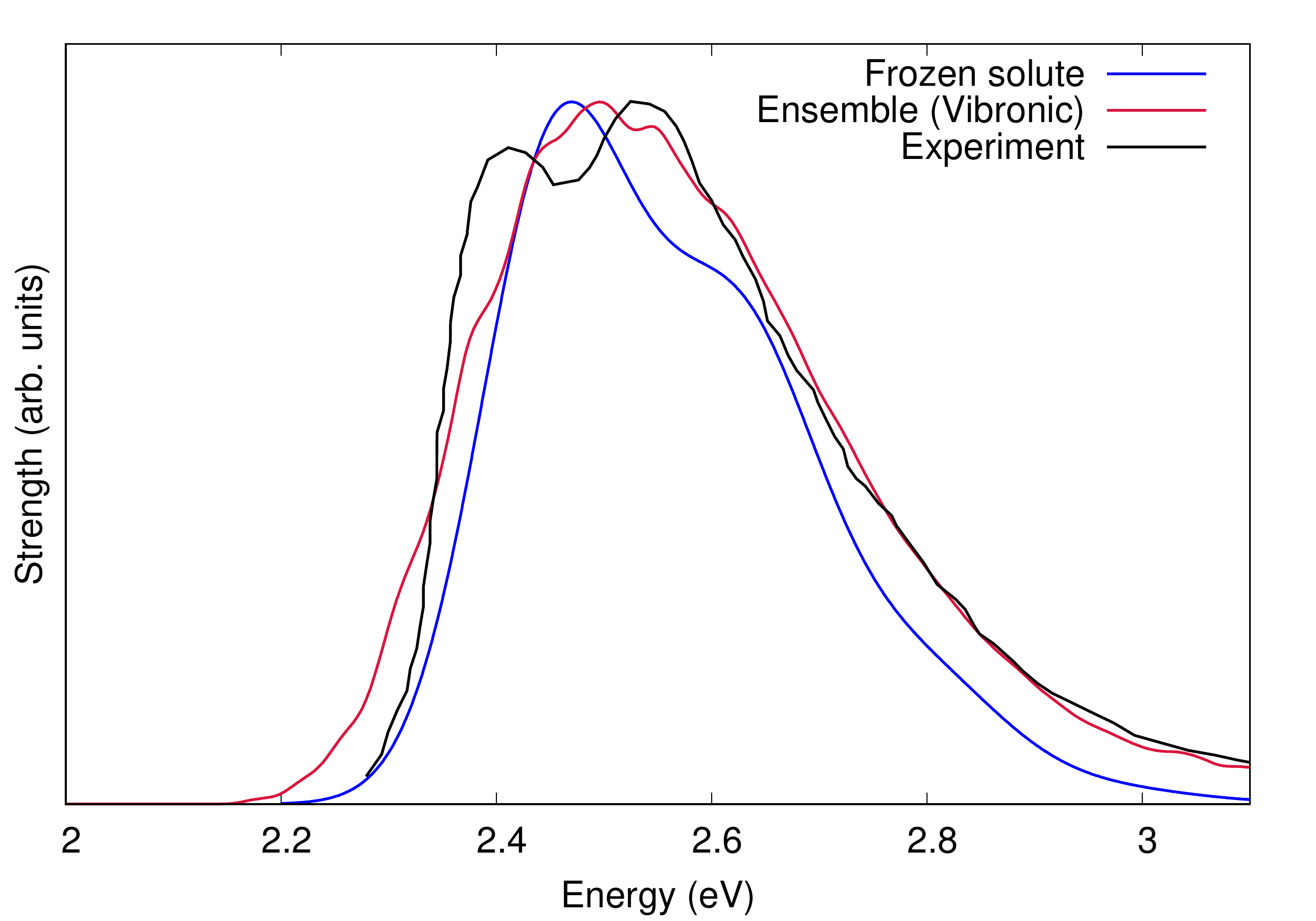}}} \\
\subfloat[Nile Red in benzene\label{subfig:nile_red_benzene2}]{\resizebox{0.45\textwidth}{!}
{\includegraphics[angle=270]{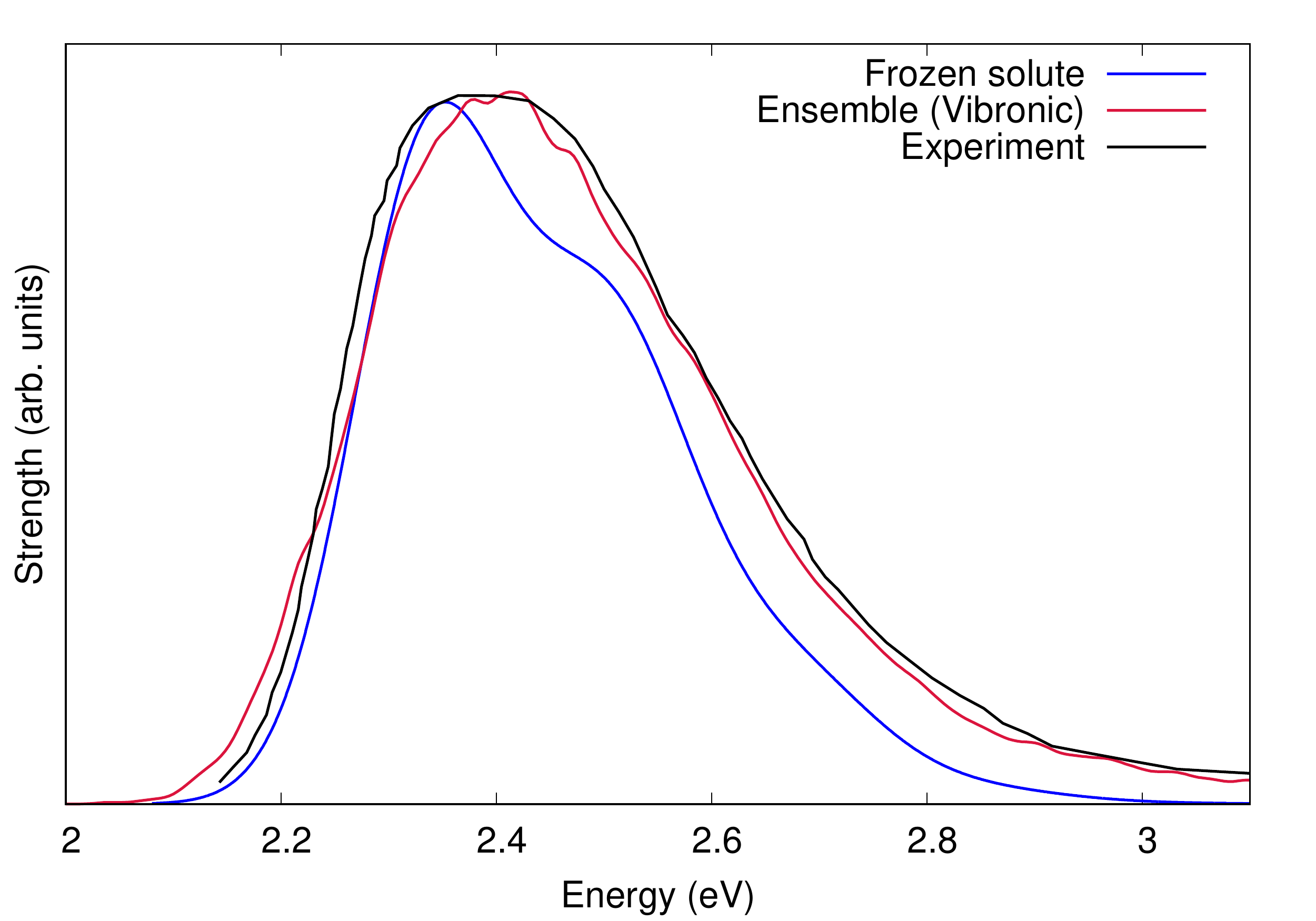}}} \\

\subfloat[Nile Red in acetone\label{subfig:nile_red_acetone2}]{\resizebox{0.45\textwidth}{!}
{\includegraphics[angle=270]{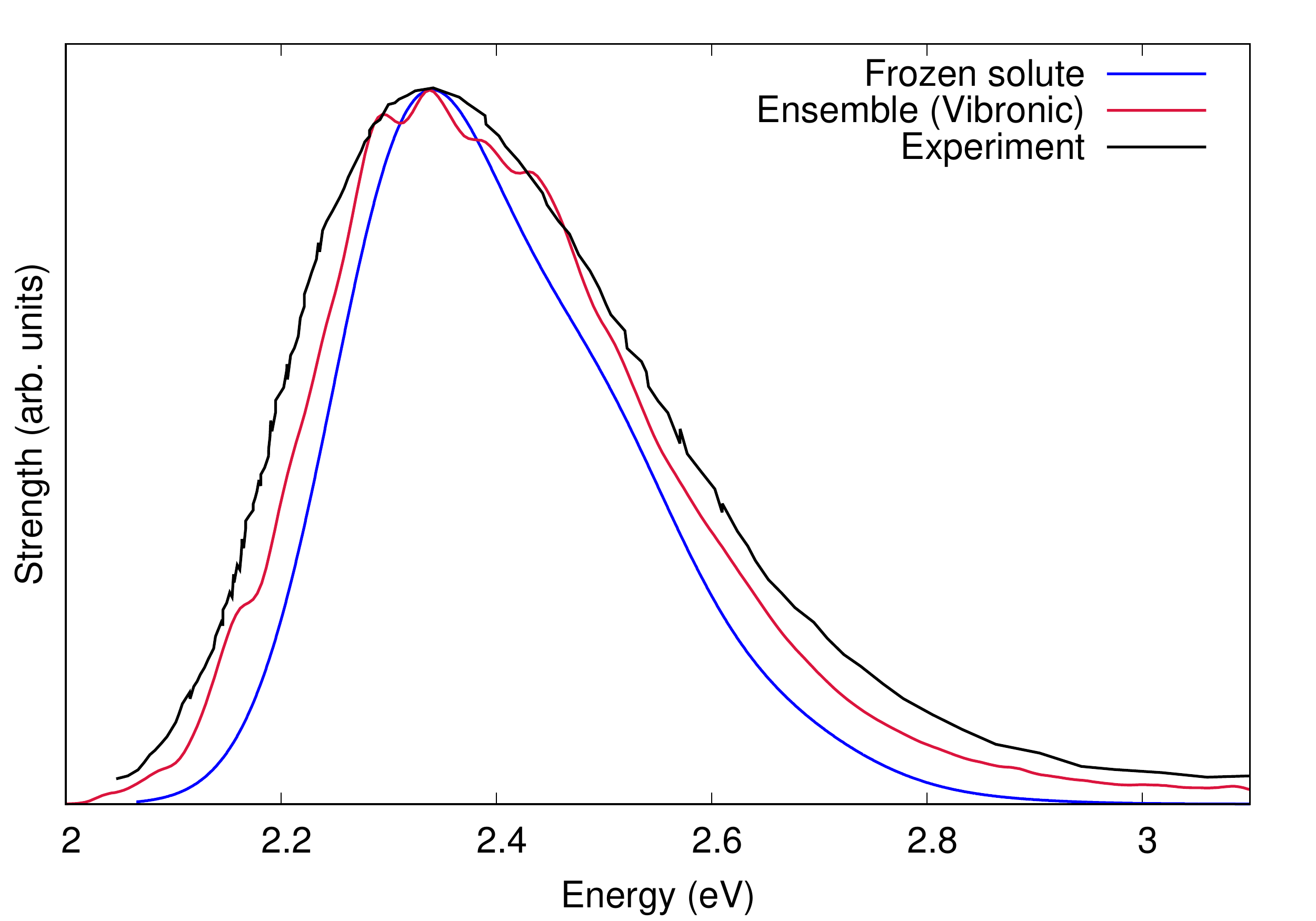}}}
\caption{Absorption spectra for Nile Red in three different solvents, as calculated with the vibronic broadening approach, as well as the frozen solute approach of Santoro \emph{et al.}\cite{Santoro_vibronic}. The Gaussian broadening factors are computed from vertical excitation energies of 500 uncorrelated MD snapshots of the frozen ground state solute structure in solution. The broadening factors computed in this way are $\sigma=0.073$ eV for acetone, $\sigma=0.066$ eV for benzene, and $\sigma=0.065$ eV for cyclohexane. All calculated spectra are shifted and scaled in order to match experimental data. }
\label{fig:nile_red_santoro}
\end{figure}

Given the comparatively weak solute-solvent interactions that are expected to occur for Nile Red in the three chosen solvents, we also test how well the experimental spectra are reproduced by the approach of Santoro \emph{et al.}\cite{Santoro_vibronic}, which relies on a complete decoupling of solute and solvent degrees of freedom. To do so, we follow an identical calculation protocol as the one described in Ref. \cite{Santoro_vibronic}. The solvent-induced inhomogeneous broadening is modeled by applying Gaussian broadening to the Franck-Condon spectra of Nile Red in implicit solvent, as computed at 300 K. The standard deviation of the appropriate Gaussian broadening is obtained by fitting to the distribution of vertical excitation energies computed from 500 solute-solvent conformations with the solute frozen in its ground state optimized structure. The standard deviations for the inhomogeneous solvent broadening computed in this way are $\sigma=0.073$ eV for acetone, $\sigma=0.066$ eV for benzene, and $\sigma=0.065$ eV for cyclohexane, respectively. 

The resulting spectra, as compared with the spectra computed from our vibronic broadening approach and the experimental spectra, can be found in Fig. \ref{fig:nile_red_santoro}. As can be seen, the spectra produced by decoupling the solute and solvent degrees of freedom are too narrow compared to experimental results, which is consistent with the results reported in Ref. \cite{Santoro_vibronic}. Also, similar to the results reported in \cite{Santoro_GFP_solution}, the broadening is not strong enough to fully wash out the the vibronic fine structure of the transition, with a shoulder present in all spectra. The change in spectral shape from a double peak structure in cyclohexane to a single peak in acetone is also not recovered correctly and the spectrum of Nile Red in benzene is wrongly predicted to be very similar to that in cyclohexane. In cyclohexane the method suffers from a very similar failure as the combined ensemble plus vibronic broadening approach introduced in this work, in that the very sharp absorption onset at 2.3 eV is not reproduced. The more gradual onset of the simulated spectrum is a direct result of applying a Gaussian broadening to the intense 0-0 transition of the Franck-Condon spectrum. We note that, unlike in the approach introduced in this work (Eqn. \ref{eqn:vib_broadening_solvent}), where a sharp absorption onset can originate if the underlying distribution of bare vertical excitation energies shows a sharp onset, the same is not true for the approach introduced in \cite{Santoro_vibronic}, since it always assumes a Gaussian distribution of the inhomogeneous solvent broadening. 

Although the bare vertical excitation energies of the unconstrained solute-solvent conformations are distributed in a very non-Gaussian way, the energies of the systems where the solute is frozen in its ground state structure are considerably more Gaussian (see Section IV of the supporting information). In fact, both acetone and cyclohexane produce inhomogeneous solvent broadening effects that are fully consistent with a Gaussian distribution, in line with the main assumption of the approach in Ref. \cite{Santoro_vibronic}. The benzene solvent environment, however, produces non-Gaussian inhomogeneous broadening. This result can be straightforwardly interpreted by considering that aromatic solvents such as toluene and benzene have been shown to form $\pi$-stacking conformations with the Nile Red chromophore\cite{Nile_red_force field}, such that the first solvation shell is relatively ordered. In this scenario, the assumption that random arrangements of solvent molecules around the frozen solute produce Gaussian noise starts breaking down. It is noteworthy that this breakdown already occurs for relatively weakly interacting solute-solvent systems, suggesting that it is less valid for protic solvents like water or methanol, as encountered in the previous section. 

The combined approach introduced in this work dresses the bare vertical excitations with an average vibronic fine structure; the method accounts for temperature fluctuations in a purely classical way through MD, but goes beyond the approach of Santoro \emph{et al}. by allowing for solvent-induced conformational changes of the solute, as well as solute-solvent coupling in the calculation of the excitation energies. Although some spectral features of the absorption spectrum of Nile Red in going from cyclohexane to acetone cannot be correctly reproduced, most notably the double peak feature in cyclohexane, this failure can potentially be traced to the conformational sampling carried out using classical force fields. The results presented here, together with the successes in reproducing the experimental spectra of the GFP chromophores in solution demonstrated in the previous section, provide good evidence that the simple approach of adding vibronic fine structure to vertical transitions via a vibronic shape function is capable of correcting the main failures of the ensemble approach.

\section{Conclusion}
In summary, we have introduced a combined ensemble plus vibronic broadening approach that provides a simple correction to absorption spectra of solvated dyes computed from vertical excitations of solute-solvent conformations. In this approach, all temperature fluctuations are treated classically from the sampling of configurations, whereas the quantum nature of the solute nuclei is approximately treated at zero temperature. In this work, the nuclear quantum effects are accounted for by a calculation using the Franck-Condon principle, i.e. by considering the overlap between the ground state vibrational mode of the electronic ground state and vibrational modes of the electronic excited state. This combined approach can be interpreted as `dressing' the bare vertical excitations with a Franck-Condon spectrum. 

Here, the Franck-Condon spectrum was computed as an average from a small number of frozen solvent conformations.  A more rigorous implementation of the combined approach would be to compute a Franck-Condon spectrum for each solute-solvent configuration, while keeping the solvent frozen, and using a larger number of explicit solvent molecules than was used here. However, this would come at a very steep computational cost with perhaps no significant improvement in overall spectral shape. 

For the small to medium-sized systems studied in this work, the combined approach presented here is not significantly more expensive than approximating an absorption spectrum from the ensemble approach alone, but corrects the spectrum to capture the vibronic asymmetry of experimental spectra. In the combined approach, both inhomogeneous solvent and temperature broadening effects are captured from first principles and no artificial smearing parameters have to be introduced to reproduce the width of experimental spectra, apart from a small, system-independent numerical broadening factor accounting for finite sampling. The method yields a strong improvement over the pure vertical spectra for all systems studied, and performs well for both systems with strong and weak solute-solvent interactions. 

The approach presented here should be considered a simple correction to spectra generated by bare vertical excitations, and thus cannot be derived from a rigorous separation of solute and solvent degrees of freedom as other approaches\cite{Santoro_vibronic}. However, no assumptions are made about functional form of the influence of the solvent degrees of freedom on excitation energies of the solute, which in this work is shown to be non-Gaussian in nature, even for systems with comparably weak solute-solvent interactions. Furthermore, although the use of a single average vibronic shape function clearly has its limits in systems with strong solute-solvent interactions, as seen in the case of the GFP anion in water, the main features of the experimental absorption spectra are still reproduced correctly. 

The simple approach suggested in this work will necessarily reach its limits for very flexible systems and systems undergoing photoisomerization, where the overlap between the ground state and excited state nuclear wavefunctions becomes small and the Franck-Condon principle breaks down. Furthermore, for very large dyes, the computation of the vibronic shape function in terms of a normal mode analysis of the ground and excited state optimized structures is likely not computationally feasible. Lastly, in systems where the quantum nature of initial vibrational states beyond the ground state becomes important, treating temperature fluctuations in a purely classical sense can no longer be valid. However, for a large class of small to medium-sized semi-rigid dyes, we expect the combined ensemble plus vibronic broadening approach presented here to produce spectral shapes in close agreement with experimental results.  

\begin{acknowledgements}
This work was supported by the Department of Energy, Office of Basic Energy Sciences CTC and CPIMS programs, under Award Number DE-SC0014437.  This work used the XStream and MERCED computational resources supported by the NSF MRI Program (Grants ACI-1429830 and ACI-1429783).
\end{acknowledgements}

\end{document}